\documentstyle[12pt,worldsci]{article}
\newcommand{\be}{\begin{eqnarray}}
\newcommand{\ee}{\end{eqnarray}}
\begin{document}

\title{THE QCD VACUUM, CHIRAL PHASE TRANSITION AND QUARK-GLUON PLASMA  }

\author{Edward Shuryak\\
{\it Physics Department, State University of New York,\\
Stony Brook NY 11794, USA\\
email shuryak@nuclear.physics.sunysb.edu}%
}
\maketitle

\section*{Abstract}
This is a brief review of three strongly related topics. In recent years,
significant progress was reached in understanding of vacuum and hadronic
structure: quantitative role of non-perturbative tunneling,
described semiclassically by
``instantons" was clarified.
 We review recent works on
the {\it point-to-point correlation
functions}, comparing those obtained from phenomenology,
 ``instanton liquid" models and
on the lattice.  The second topic
 is physics of chiral symmetry restoration, which may lead to
observable hadron modification and unusual event-per-event fluctuations.
We discuss collective flow and phenomena related with
a remarkable ``softness" of the equation of state near the phase transition,
as well as phenomena at small $p_t$.
 We also describe recent ideas on the mechanism of chiral restoration, based on
  formation of {\it polarized instanton-antiinstanton molecules}.
The third part, related to quark-gluon plasma, is
related mainly with the issue of initial equilibration of
QGP in high energy collisions. We discuss different ``parton cascade"
approaches and argue that
{\it multi-gluon} processes dominate
them. We discuss  predictions for  RHIC and LHC energies.

\section{Introduction}

   Studies of high-temperature (high-density) hadronic
matter has grown into a wast field during the last decade.
 In  70's there were just few
theorists dreaming about  phase transitions in QCD, and now we have
conferences with hundreds of participants, most of them experimentalists.
If in early 80's all attempts to preserve ISR from destruction failed, now the
nuclear physics community in US has united and pushed forward
 a dedicated complex, RHIC, now under construction. In Europe, heavy-ion
experiments are planned at the largest accelerator to be made at CERN, LHC.

   When our collegues from other
fields ask for a brief explanation of the major motivations of this program,
we usually say that high temperature matter is relevant for Big Bang, remind
them that high
density exist in compact stars, and conclude that
in quark-gluon plasma is a new state
of matter and it is interesting in its own rights. It is all true, but it is
far
from being all of that.

   A very important part of it is related with hopes to understand better the
world we are living in. We know that QCD is the fundamental theory
of strong interactions, but its ground state, {\it the QCD vacuum },
is not understood. It is important to realize that it is in fact a very
complicated and  quite dense matter by itself.
When all perturbative infinities are taken care of,
one is left with the so called {\it non-perturbative energy density}
$\epsilon_{vac} \approx -1 GeV/fm^3$. The minus sign is important: it means
that
the physical vacuum is below the naive ``normal" one, in which only small
zero point fluctuations take place. So to say, we all live in a kind of a
superconductor\footnote{
According to Standard model, Higgs fields create its own
version of a superconductor, which can also be ``melted" at $T_c$ about a
thousand times that for QCD: but we would not touch the electroweak physics
in this review.   }, and only by producing a tiny fireball of QGP we can
learn about existence of this ``normal" phase. Clearly, even such
crude information as
the value of the critical temperature $T_c$ is very important
to understand of the phenomena underlying this energy density and excitations
in our world. By studying how how hadrons ``melt"
we may learn more about their structure.

  Large value of the vacuum energy density explains why we need really high
energy accelerators: we have to melt it first, before filling the space with
thermal quarks and gluons.  And only when we will be able to do so, the
physical reality of this ``vacuum energy" (and the ``vacuum pressure"
$p_{vac}=-\epsilon_{vac}$ ) will become obvious. (This situation resembles
very much what have happened when people has realised the reality of
atmospheric pressure.)

  But where this big negative energy comes from? An answer is not completely
clear yet, but  I will argue below that it comes from {\it tunneling}
phenomena, between certain topologically different configurations of the glue.
(As we know from quantum mechanics, if one has several potential wells
with a bound state, tunneling from one to another does lower the ground state
energy.) Of course, one would
 need a truly quantitative theory of those phenomena and a
confirmation by the experiment in order to be quite sure about it.
So far, we are at least sure that tunneling  contribution to the vacuum energy
shift has  right order of
magnitude.

   In chapter 2 of this review I will describe
a dramatic progress which took took place during the last 2-3 years
 which has resulted in
 $ quantitative$ understanding of tunneling
phenomena in QCD. We knew since mid-70's, that one can use semiclassical theory
based on the ``instanton" solution to describe them,
provided the action along the tunneling path is much larger than the Plank
constant $S>>1$. Unfortunately,
it is not generally clear whether it is the case, or other competing tunneling
paths take over.
A real
breakthrough was related with studies of the QCD correlation functions
(see ref\cite{Shuryak_cor} for a review), which
has demonstrated quite clearly that instanton-related forces between quarks are
indeed there and are indeed very very important. To mention one of those (quite
unexpected) findings:  the $nucleons$ are
actually bound mostly by those forces, not by a confining ones, as we thought
before. Furthermore, the nucleon-delta (spin) splitting also seems to be
mainly an
instanton effect, not the gluo-magnetic spin interactions
 we used to believe in.

    At this point of the introduction, let me make some
    general remark on  our
most powerful theoretical
tool, the lattice gauge theory.
In the last couple of years we are witnessing now
a qualitative new stage of its development.
Ten years ago, the main concern in the field was whether the results
make sense: people have checked universality, scaling, etc. Few years ago
lattice community became convinced that it may not only reproduce
experimentally known
masses and other parameters (still with much worse accuracy), but go ahead
and make some quantitative predictions of unknown parameters (e.g.
$\Lambda_{QCD}$.)
  A new step are   attempts to identify
{\it the most important  structures}
in a very complicated vacuum.
   Two types of very interesting structures were identified so far: (i)
{\it instantons} and (ii) {\it paths of the monopoles}.
 In the first case,
it was the so called ``cooling" (see details in
ref. \cite{Negele}), in the second one it is the so called ``abelian
projections" \cite{Suzuki}.
In both cases
  a kind of ``radical surgery" was used,
eliminating all fields except of those of an interesting structure. And
in both cases it was demonstrated that the main phenomenon under
consideration
(chiral physics and confinement, respectively)
do survive this operation.
   What should be strongly encouraged at this point,
is a continuation of these efforts, and their generalization for non-zero
temperatures (and densities, if the working method of doing it will be found).

   The next chapter  is related with chiral
phase transitions and its possible manifestations in experiment.
In subsection 3.1 we discuss the fate of $two$ chiral symmetries, related with
$SU(N_f)$ and U(1) groups. Then we consider possible experimental approaches
to search for signatures of the phase transition.
 Those
potentially include the whole range of ideas, from simple utilization of
the ``softness" of equation of state and flow, to mass and width
modification of hadronic modes, and eventually to very complicated
questions related with critical fluctuations. We also consider
new ideas, relating chiral restoration to formation of instanton-antiinstanton
molecules.

   In the last chapter we briefly address some topics related with QGP itself.
We do not consider neither the progress in partial resummation
of perturbation theory, nor those concerning
the non-perturbative
phenomena at very-high T. Instead, we have concentrated on QGP production and
equilibration,
at RHIC/LHC energies.
We compare different theoretical tools used, as well as some of the results.

\section{Correlators in the QCD vacuum and instantons}

\subsection{Vacuum and hadronic structure}

   Let us start with
   brief recollection of the history of hadronic physics.
After many hadronic states were discovered in 50's,
it became clear  that they should be some composite objects.
In 1964, we have learned from Gell-Mann and Zweig
that hadrons are made of quarks.
The first model on the market was (i) {\it constituent quark model}, in which
a nucleon was viewed as a
non-relativistic bound state
of three separate $massive$ quarks. It
 works surprisingly
well in some cases, e.g. for baryon magnetic moments. However,
as it became obvious already by the end of 60's, ``true"
 quarks are actually nearly massless\footnote{
Here and below we discuss only hadrons made of light (u,d,s)
quarks. Of course, things are very different in the world of heavy
quarks: say $J/\psi$ and $\Upsilon$ physics is well described by the
non-relativistic potential, a combination of a Coulomb and confining ones.
}
, so that the chiral symmetry is nearly
exact. The first part of this idea (light quarks) were well incorporated into
the favorite model of 70's, (ii) {\it the MIT bag model}.
 The second part (chiral
symmetry) was badly violated in this model, and
therefore  a completely
different model for a nucleon became popular, in 80's,
known as  (iii) {\it the skyrmion}, in
which nucleon is entirely made of a pion field.
Later, another model
was suggested, connecting the bag
and the chiral symmetry together, into (iv) {\it the chiral bag model}.
   Looking at  these models together, one could not be surprised
that our students are confused: these models
suggest drastically different  pictures of hadronic structure. While
the MIT bag model relates all dimensional quantities
to a ``bag
constant", representing a confining force,
the skyrmion picture
has no place for confining forces at all!

  Such situation clearly reflects the fact, that this field is still
not mature enough, in spite of its age. All
 those models cannot be
true, and one has to work harder to tell which ones are wrong.
My point here is that the information used so far
(contained in the spectra
of the low-lying states) is simply insufficient to make this choice.
The same was true for nuclear forces: deutron data are important, but only
complete set of scattering phases has clarified them in sufficient details.

  Unfortunately, one cannot do  quark-antiquark scattering: but
one is still
able to connect the
hadronic structure to the fundamental theory,  QCD, by calculating
Euclidean correlation functions for all distances and all channels.
Hopefully, this will be enough.

   My second general remark: all models of hadronic structure mentioned above
ignore one general principle, which we have learned solving numerous problems
of quantum physics. This principle is: {\it solve the ground state first, then
it is easier to understand the excitations}.
   All these models try to avoid the question about the {\it vacuum} structure,
in one way or another. The MIT bag model, for example, does
acknowledge existence
of non-perturbative effects in QCD, but only {\it outside} the hadrons.

    In fact, in the list of models mentioned above
 one important model of hadronic structure (actually, the oldest
one) was missing. This is the Nambu-Jona-Lasinio \cite{NJL}
(below NJL) model, which was inspired by the BCS theory of superconductivity.
This one actually has the right strategy: {\it assuming} existence of some
short-range
attraction
between quark and antiquark
(in the present-day notations),
 and moves first to the discussion of the ground state.

   Let me briefly sketch here a picture of light-quark hadrons, as I see it
today.
Ironically enough, Nambu-Jona -Lasinio
model is actually right
in assuming that
strong short-range attraction is the main effect. It does exist in QCD,
 generated by the tunneling phenomena we are going to discuss below\footnote{
Although the exact effective interaction
is not exactly the same, of course, and even have different symmetries.}
Those forces make pion light, $\eta'$ heavy and build the quark condensate.
``Constituent quarks" are relatively heavy, with $M>400 MeV$, while some
hadrons (in particularly, the nucleon) are deeply bound. Even $\rho$ and
$\Delta$ baryon  probably remain (weakly) bound,
 if one switches off confining forces.

\subsection{Instanton history in brief}
    Tunneling phenomena in gauge theories, discovered
by Polyakov
and collaborators \cite{Belavin_etal_1975},
 were soon followed by the fascinating
semiclassical treatment
by 't Hooft \cite{tHooft}. It was pointed out, that tunneling  lead to
completely new type of effective interaction between light quarks.
This interaction actually
 explains how chiral anomalies work. The
first applications to QCD problems such as ref.
 \cite{CALLAN_DASHEN_GROSS_1978}
 have attracted a lot of attention in late 70's.
 However, as no explanation
for ``diluteness"  and  validity
of semiclassical approximation
were suggested from the first principles, optimism has soon died out and
 most people  left the field.

  Next  period has mainly focused on
 phenomenological manifestations of instanton-induced effects.
Starting from QCD sum rules and its problems,
the so called  ``instanton   liquid model"
 \cite{Shuryak_1982} has emerged. It
suggested relative diluteness and large action per instanton due to
their relatively small size, but also emphasized significant interaction
in the ensemble as the origin of density stabilization.
   Attempts to describe {\it interacting} instantons
were initiated by the {\it variational approach}
 \cite{DIAKONOV_PETROV_1984}. For the simplest ``sum ansatz"
for the gauge fields it was shown that there appears
repulsive interaction at small
distances, which may stabilize the density and lead to
qualitatively
  correct instanton liquid.
Further numerical studies of this problem \cite{Shuryak_1988}
have allowed to get rid of many approximations and
eventually included fermionic effects
to all orders in {\it 't Hooft effective Lagrangian} \cite{tHooft}.
We return to  discussion of this approach below, and now let me jump directly
to several important steps made during the last
years.

  About 40 correlation
functions were calculated  in the framework of the simplest
ensemble of the kind, the Random Instanton Liquid Model (RILM)
 \cite{SSV_cor}. Agreement with data is
generally good, and
in some cases (including $\pi$,N etc) it is really astonishing.
Recently, glueballs were added to the list \cite{SS_glue},
and therefore the instanton liquid is certainly by far the best model
 of hadronic and vacuum structure, available
today.

  On the theoretical side, howvere,
  many important questions are left open, though.
 Is the classic repulsive interaction
really there, or another explanation (e.g. related to confinement or charge
renormalization) is need for density stabilization?
  Is this ``random liquid" really a reasonable
approximation? What is the role of interaction
between instantons, especially of quark-related one? What happens for larger
number of colors, or flavors?

  Few words about instantons on the lattice. First of all,
  lattice calculation of point-to-point correlation functions
 \cite{Negele1} also show good
agreement with experiment \cite{Shuryak_cor} and  the RILM results
 \cite{SSV_cor}.
 Furthermore, experimentation
    with  ``cooled" lattice configurations \cite{Negele}
has essentially confirmed parameters of the original ``instanton liquid" model.
 However, a lot of work
is still needed. In particularly,
both lattice and instanton studies mentioned are some sense
``quenched" so far, and inclusion of dynamical  quarks
should be done. This is very important at finite temperature,
especially near the QCD phase transition.

 Finally,
about a separate brunch of the instanton studies.
Few years ago the instanton-induced $processes$  has attracted much attention
after works by Ringwald, Espinosa and others about
baryon number
violation in weak interactions.
A hope was expressed for some time, that it would be possible to observe it
experimentally, in multi-TeV collisions, but it looks that the relevant
cross section
are orders of magnitude below the level reachable by any
 experiments. However, ideas developed in this
context may be used in the QCD context.
  Interesting examples are the {\it instanton-induced}
deep-inelastic scattering, or jet production, see more in refs\cite{BB}.

\subsection{Why instantons?}

   We have already mentioned that the main
reason why instantons are so important for
physics of {\it light fermions}: it is related to the
$anomaly$ phenomenon, which means that
 each tunneling event leads to rearrangement of quark
chiralities. Since that happens for all three light quarks u,d,s at the same
time, one gets generally a 6-fermion effective interaction.

  Let us however change the language, and outline another possible approach,
which is the one actually used in calculations. Anomalies are related to
 famous 't Hooft
zero modes, the localized solutions of the Dirac equation
\be  D_\mu\gamma_\mu \phi_0(x)=0 \ee
where D is the covariant derivative containing
the instanton field. Evaluating
the (Euclidean)  quark propagator
$ S= - 1 /[ iD_\mu\gamma_\mu +im] $
for $m \rightarrow 0$ and large distances
 one has to deal mainly with  small eigenvalues.
The fermionic states which are ``made out of" zero modes naturally have this
property, and thus are ``prime susspect" for being the relevant ones.

  It is therefore
convenient to look at the instanton as a $trap$ for quarks, something
 like a ``receptor atom" in
a semiconductor, capable to create a new state in otherwise forbidden place.
 In metals an electron
can propagate far, just by hopping from one atom to another. The same is
true for finite instanton density, leading to a ``zero mode zone"
 of collectivize quark states. That is the
 mechanism leading to the non-zero quark condensate, or chiral symmetry
breaking.

   Of course, in order to make these statements completely convincing, one
should be able to check whether the ``zero mode zone" does indeed
dominate in
the
density of states at ``virtuality" $\lambda=0$.
 The QCD vacuum has a lot of different fluctuations of the color
fields (e.g., the monopole loops):
 and one may imagine that most of them may contribute somehow to
it. This can be done by tracing the nature of all lowest-$\lambda$ fermionic
states.

   Furthermore, if more than one quark is travelling
in the
 QCD vacuum ($\bar q q$ for mesons and qqq for baryons), they  ``hop"
over the same instantons. This fact leads to an effective interaction, which,
if attractive enough, may
 in fact  bind quarks together.

   We have claimed that instantons are more important than
 any other fluctuations
of the gauge field. To prove that phenomenologically,
 one has to look more specifically into
the chiral and flavor structure of the instanton-induced interaction.
As shown by 't Hooft,
 at tunneling quarks with one chirality
 ``dive into the Dirac sea" while those  with the opposite chirality
 ``emerge" from it. Therefore instanton-induced forces should be
strong in {\it scalar} and {\it pseudoscalar} channels, and absent (in first
order) in vector or axial ones.
Looking at phenomenological correlators at small distances, one finds that it
is exactly right.

   Furthermore, consider signs of the instanton-induced corrections to
spin-zero channels. The correlation functions
at small distances are essentially
the free propagators squared (for mesons, and cubed for baryons).
 If the instanton-induced corrections
are relatively small, one may use the
't Hooft interaction in the lowest order.
 There are 4 such channels for 2 flavors
and it is a simple matter to see that correction is positive (or {\it
attractive}) for $\pi,\sigma$ channels and negative (or {\it repulsive})
for $\delta,\eta'$ ones.

Thus {\it the same}\footnote{   Note at this point, that many other models
(e.g. those based on confining forces) for chiral symmetry breaking
lead to light pions as well: Goldstone theorem simply demands it.
However, for those models there is no hope to get the $\eta'$ right.
} mechanism leads to
both light pion and heavy $\eta'$! Both splitting from a typical
meson like $\rho$ are  large, which is
a very strong hint. Moreover, looking at masses $m_\pi^2\approx 0,
m_\rho^2\approx 0.5 GeV, m_{\eta'}^2\approx 1 GeV$ one finds the splittings
to be even comparable!

\subsection{Correlators in the instanton vacuum}

   Let me recall here  very
deep intuitive thoughts by Yukawa:
 in order to recognize existence of a particle,
one should $not$ necessarily find it in the detector, or observe a pole
in a  propagator. In fact one may just observe an amplitude,
 exponentially decaying
with $distance$ to recognize existence of a {\it virtual} particle.
This is exactly how all hadronic masses are measured nowadays on the lattice.

  One may exploit this idea further and consider
a correlation of two local operators, separated by the space-like
distance r. For example, the operators can create a quark-antiquark pair.
At  not-very-large r, the correlator is not falling exponentially with a single
mass, and it cannot be described in terms of {\it one}
 propagating meson: but it
still provides a lot of information about the
 quark-antiquark interaction.
If such correlator is known,
it allows us to tell the
effect of the {\it  short-range forces} (e.g. instanton-induced ones, to be
much discussed below) from the {\it long-range}
 ones (e.g. confinement-related).

  Now we proceed from  qualitative hints to quantitative calculations.
We have to
 evaluate
 a quark propagator in the multi-instanton
field configuration, which can be done
 as follows:
\be S(x,y)=\Sigma_{ZMZ}
 {\phi_\lambda(x) \phi^+_\lambda(y) \over \lambda -im }+ iS_{NZM}(x,y))
\ee
where the first term is the sum over states belonging to the ``zero mode zone".
The non-zero modes
(analogs of unbound atomic states) are taken into account by the last term, see
details in
\cite{SSV_cor}.

  We have first calculated correlators for the
  simplest ensemble possible, the {\it random instanton liquid model}
 (RILM), in which: (i) all instantons
have the same size $\rho_0=.35 fm$; (ii) they have {\it random}
 positions and orientations;
(iii) instanton and anti-instanton densities are equal, and in sum it is
$n_0=1 fm^{-4}$.
  These are the parameters suggested a decade ago in \cite{Shuryak_1982}, the
density comes from
the gluon condensate and size from various other things, say from
the quark condensate value.
The main step in the calculation is inversion of the Dirac
operator, written in the zero-mode subspace. (We typically
use in sum 256 instantons and
 anti-instantons, which tells the dimension of this matrix and the volume of
the
box.)

   Although the quark propagators are gauge dependent, we have looked
at them first in order to see whether
they can be reproduced by any simple model, say by ``constituent quarks" with
a constant mass. We have found that chirality-non-flipping part of the
propagator
indeed looks as if quark get a mass about 300-400 MeV, but the
 chirality-flipping part  does not look like that
at all. None of many calculated correlators in fact follow a constituent quark
model.

\begin{figure}[t]
   \vspace{6.in}
   \caption{ Correlators for pseudoscalar (P) and vector (V) channels
 according to
RILM  (open points)
and lattice results
(closed points) versus distance x in fm. The correlators are normalized to
those
corresponding to free quark propagation: therefore at small x all of them tend
to 1. The long-dash lines correspond to experimental data (other lines
are used for fitting of lattice data).
}
\end{figure}
\begin{figure}[t]
\vspace{3.5in}
\caption{ Similar to what is shown in Fig.1, but now for N,$\Delta$ channels.
Triangles are RILM, squares are lattice results.
The lines are predictions by two works based on the QCD sum rules,
by Belyaev and Ioffe (long-dashed) and Farrar et al (short-dashed). }
\end{figure}

   Our results for $\pi,\rho,N,\Delta$ channels are shown in
Fig.1,2.
 All correlators are plotted in a normalized way, divided by
 those corresponding to {\it free} quark propagation:
that is why all of them converge to 1 at small distances.
Solid lines correspond to experiment \cite{Shuryak_cor},
while the  long-dashed and short-dashed curves correspond to
QCD sum rule predictions
\cite{Belyaev_Ioffe} and \cite{FZOZ}, respectively.

The agreement for the pion curve is as  perfect as it can be:
 both the mass (142$\pm$12 MeV) and the (pseudoscalar)
  coupling are reproduced correctly, inside
the error bars! Large deviations from perturbative behavior
happens at very small distances for the $\pi$ channel, while exactly the
opposite is observed in the $\rho$ case,
the plotted ratio
remains close to 1 up to a very large x.
Both RILM and lattice has reproduced that non-trivial observation.

  Proceeding to baryonic channels, let me mention that
 we have actually
measured all 6 nucleon correlators and 4 delta ones, and have fitted them all.
  Again, agreement between RILM and lattice results is surprisingly good,
literally inside the error bars.
Both display a qualitative difference
between the nucleon and the delta correlators: this can be
traced to {\it attractive}
 instanton-induces forces for the spin-isospin-zero diquarks.
Without any one-gluon exchange the RILM
predicts the $N-\Delta$ splitting (actually,
we have found that in RILM $m_N=960\pm 30 MeV$ and  $m_\Delta= 1440\pm70 MeV$,
so the splitting is
in fact somewhat too large).

  There is  no place here to discuss other channels in details. Let me only
mention here that
the most difficult case proved to be
 the isosinglet scalar $\sigma$, for which
 one should not only evaluate
the double quark loop term, but also subtract the disconnected
$|<\bar q q>|^2$ part. Curiously enough, we have found dominance of
a light state, with
m$\sim$ 500 MeV, reminiscent of the sigma meson of 60's.
For several  reasons such measurements are now beyond the reach
of lattice calculations, although existence of attractive interaction
at $x\sim 1/2 fm$ can probably be seen.

  Let me also mention recent studies of the
 the so called ``wave functions"
 (known also as {\it Bethe-Salpeter amplitudes}), also done for both
RILM and (quenched) lattice simulations.
The main qualitative features (e.g.: $\pi,N$ are {\it more compact} than
$\rho,\Delta$) are also reproduced.
They have shown with even greater clarity, that instantons
do lead to  quark binding in most channels involved, even without
 confining forces.
The $shape$ of the wave function is however not the same.

  Clearly, RILM discussed above cannot be but a crude
approximation:
 at least, the very phenomenon studied
above, the quark's hopping  from one instanton to another, should
lead to strong
correlation between them. Another obvious source of interaction is the
non-linear gluonic Lagrangian: a superposition of instantons and
anti-instantons
 have the
action different from the sum of the actions.
Furthermore, at least in two ``repulsive" channels (pseudoscalar
 isoscalar $\eta'$
and scalar isovector $\delta$, or $a_0$)
RILM leads to phenomenologically unacceptable results, showing too strong
repulsion induced by an instanton. Only the presence of nearby anti-instantons
may help, and this is precisely what ``unquenching" of quarks is suppose to do.

Generally speaking, the ensemble of instantons
   should be described by a partition function of the type
\be Z= \int  d\Omega  exp(-S_{glue}) [det(i\hat D
+im)]^{N_f}
\ee
(where
$ d\Omega$
 is  the measure
in space of {\it collective coordinates}, 12 per instanton).

It is
a problem similar to those traditionally studied in statistical mechanics,
with the main complication being the non-local fermionic
determinant.
  As
shown in
\cite{DIAKONOV_PETROV_1984,Shuryak_1988},
if the fermionic determinant is calculated in the ``zero mode zone" subset
of fermionic states, it includes all diagrams with 't Hooft effective
interactions. For a simulations involving N/2 instantons and N/2
anti-instantons, one has to deal with a  $N * N$ matrix.
However,  it is
still orders and orders of magnitude simpler than the lattice gauge theory!

  The simulation done already in 80's have shown
this statistical sum describes a liquid, in which
chiral symmetry {\it is broken}.
   Recent studies of correlation functions with {\it interacting ensemble}
have shown other
significant improvement over RILM.
In particular, in  recent paper \cite{SV_screening} it was shown
  how the global
fluctuations of the topological charge are $screened$ in the
 $m\rightarrow 0$ (chiral) limit.
In more practical terms, it have fixed incorrect behaviour observed in
RILM for the $\eta'$ channel.

\subsection{Glueballs and instantons}

 We have argued above,
that the bulk of hadronic physics, including chiral symmetry breaking and
properties of all major mesons and baryons can be reproduced using even the
simplest instanton ensemble, the RILM.
  In this section we consider new development concerning {\it glueballs},
based on the recent work \cite{SS_glue}.

 Let us first briefly summarize what is known about glueballs. Experimental
evidences are too uncertain and subtle to be discussed here.
The
large-scale lattice efforts are still
needed to get reliable results, but a few statements seem to be however
 established:
(i)  The lightest glueball is the scalar, and its mass is in the
     1.6-1.8 GeV range;
(ii) The tensor glueball is significantly heavier $m_{2^{++}}
     /m_{0^{++}}=1.4$,\cite{CSV*_94,FL_92} with the $0^{--}$  probably
     heavier still.
(iii)The sizes of scalar and tensor glueballs were found to be
     drastically different. This can be inferred from  the
     different magnitude of finite size effects (see e.g. \cite{CSV*_94}),
     or seen  directly in glueball wave functions
     \cite{isst,FL_92}. The scalar glueball seems to be very compact, with a
size
(to be defined below)
     $r_{0^{++}}\simeq 0.2 fm$, while the tensor is huge with
     $r_{2^{++}}\simeq 0.8 fm$.
Clearly, this picture is very different from naive expectations, and such
drastic difference between the glueballs cannot come from confining forces
alone.

   In \cite{SS_glue}
 we propose an explanation to these phenomena based on small-size instantons.
 The main point here is that the QCD vacuum contains
small spots of very strong gluon fields, with a specific chirality structure.

As a result, relatively heavy states, with specific spin splittings, are
produced. Specifically,
instantons generate
{\it attraction} in the scalar channel, {\it repulsion} in the pseudoscalar
one and {\it no interaction} in the tensor case. (The last case is a
consequence
of the fact that the stress tensor  of the selfdual field of the
instanton is zero.)
The splittings are much stronger than in normal mesons,
because
large instanton action $S_0=8\pi^2/g(\rho)\sim 10$ enters (quadratically) here.
(For mesons, the role of classical field is played by fermionic zero modes,
which are however normalized to 1 instead of to the action.)

The results
for correlation functions are shown in Fig.3. Note that scalar channel is not
changed much if
dynamical quarks are included, while the pseudoscalar one does,
showing strong $\eta'$ signal. Our fitted glueball mass is around 1.6 GeV, and
the threshold in the pseudoscalar channel is approximately 3 GeV.

\begin{figure}[t]
\vspace{6.in}
\caption{
Point-to-point scalar, pseudoscalar and tensor
glueball correlation functions (normalized to the corresponding
free correlators) versus distance $\tau$, in fm. Stars, triangles and
squares show results for random ensemble, that with gluon interactions,
and full ``unquenched" QCD interaction between instantons, respectively.
Solid lines are parametrization used to extract masses and other parameters,
the dashed lines are the one-instanton contribution.
}
\end{figure}
  We have also determined glueball Bethe Salpeter amplitudes
(or `wave functions').
The scalar one is indeed found to be strongly decreasing function:
it can be described by $\exp(-\delta/R)$ with $R= 0.2$ fm. The
tensor wave functions has a much larger size, $R\approx $ fm.
   Further studies of the decay modes of these glueballs, both on the lattice
and in the instanton models. The suggested hierarchy of sizes,
from very small scalar to large tensor,  is of course of great relevance to
phenomenological efforts to locate these states among the observed candidates.

\section{Chiral symmetry restoration}

\subsection{Phase transitions and two chiral symmetries}

   The QCD  undergoes a phase transition at high temperatures,
to the so called quark-gluon plasma phase. Let me present at Fig.4
 a sample of recent
data \cite{DeTar} from MILC lattice collaboration. One can see, that
the transition is very rapid, the energy density is rising very rapidly
in a narrow region, of only few MeV width.
\begin{figure}[t]
   \vspace{5in}
   \caption{ Equation of state for  lattice simulations with 2
light quark flavors, from MILC
collaboration. Upper and lower points are for energy density and pressure,
respectively.}
\end{figure}

\begin{figure}[t]
   \vspace{3.5in}
   \caption{Schematic phase diagram of QCD as a function of quark masses,
from Columbia group.
}
\end{figure}

 More specifically,
since this theory have some continuous parameters, the quark masses
$m_u,m_d, m_s$ (we can safely ignore charm and heavier flavors), and thus
the phase diagram can be plotted in the corresponding 3-dimensional space.
Taking $m_u=m_d$ one gets a plot \cite{Christ} shown in Fig.5 .
It was found  from lattice simulations, that on this diagram
there exist {\it two} (seemingly disconnected) regions of strong first order
transitions: one includes pure gluodynamics (all masses very large), and
another including the point at which all masses are zero. By tradition,
the
former one is referred to as the {\it deconfinement} transition, and the latter
one as {\it chiral symmetry restoration}. Of course, these first order
transitions are separated by lines at which the transition is second order.
Other second order transitions are expected on `sides', when only one of the
masses is nonzero.

  One major physical question remains open. First, calculations with
Kogut-Susskind fermions were done by a Columbia group \cite{Christ}, indicated
that
the critical strange quark mass is rather small and therefore we should not
have a phase transition in the real world. However, recent results obtained
with Wilson fermions \cite{Kanaya} give the opposite result: even for
$m_s \simeq $
400 MeV with $m_u = m_d \simeq 0$ clear two state signals are
observed, suggesting a first order QCD phase transition in the
real world. Clearly we have to wait for few years till next generation
of simulations with dynamical fermions will clarify the situation.

   I have already mention a ``hunt" for abelian-projected monopoles,
as a source of confinement. Naturally, it was checked whether deconfinement
transition can be explained
by those objects. In ref.\cite{Suzuki} it was shown that
 the longest monopole loop (which is responsible
for the string tension) indeed disappears in the deep deconfinement region.
Furthermore, its behaviour reproduces T-dependence of the string tension.
(At the same time, the so called ``spatial string tension" remains non-zero
even at high T, as it should.)

   Before we concentrate below on the chiral transition
   let me add, that there is no such thing as the
impenetrable barrier between deconfinement and chiral restoration.
Whatever is the order of the phase transition, and whatever name we give
to it, for {\it any quark masses} there exist a rather narrow transition region
$\Delta T << T_c$ in which the energy density changes rapidly, from small value
characteristic to few hadronic degrees of freedom to a large one, ascribed to
quasi-free quarks and gluons. We should understand why it is so. No doubt,
both instantons and monopole loops significantly
change their properties in this region. Thus,
 one can easily
predict, that lattice people will find a lot of interesting things in this area
in the next few years.

For simplicity,
we ignore all effects due to the non-zero quark masses,
and consider QCD in the chiral limit, with
$N_f$ {\it massless} quarks. In this case the QCD Lagrangian is just
a sum of two separate
terms, including  right- and left-handed quarks, which implies {\it two}
chiral symmetries: $SU(N_f)_A$ and $U(1)_A$.

 Their fate is well known to be different.
 The former one is $spontaneously$
broken in the QCD vacuum  but it is restored at high temperatures, above
some critical point, denoted as $T=T_c$.
  The $U(1)_A$ chiral symmetry is not related to Goldstone bosons (as Weinberg
has first pointed out)  because
this symmetry simply {\it does not exist at quantum level}, being violated by
the 'chiral anomaly' and instantons
\cite{tHooft}.

However,
at $high$ temperatures the instanton-induced amplitudes are suppressed
due to the Debye-type screening \cite{Shuryak_conf,PY},
and therefore (at some accuracy level)
we expect this symmetry to be 'practically restored'
at high T. Let us denote the point where it happens
with some $reasonable$ $accuracy$  as $T_{U(1)}$.
  The question to be discussed now  (see more in \cite{Shuryak_pt})
is the
interrelation of the two temperatures, $T_c$ and $T_{U(1)}$. Let us
refer  as 'scenario 1' to  the case $T_c \ll T_{U(1)}$ in which the
complete $U(N_f)_A$ chiral symmetry
is restored only well inside the quark-gluon plasma domain. Another
possible case\footnote{The case $T_c >> T_{U(1)}$ does not seem to be
possible.}
$T_c \approx T_{U(1)}$ which
implies significant changes  in many hadronic channels
around this phase transition
point. As we will discuss below, these two scenarios
lead to quite different predictions.

 Pisarski and Wilczek \cite{PW}
have considered
 this question in connection with the order of the chiral phase transition.
 They have pointed out
that in the special case $N_f=2$ the
 'scenario 1' is likely to
 lead to  {\it the second order} transition. The reason is
an effective Lagrangian describing the softest modes is
essentially the Gell-Mann-Levy sigma model, same as for the O(4) spin systems.
The most straightforward way to test these ideas is to compare
the critical behaviour in both cases, testing whether
the $N_f=2$
QCD and the O(4) spin system do or do not belong to
the same universality class.

  The first critical index to compare is the one for the order parameter, for
which
the analogy \cite{RW} suggests
\be <\bar \psi \psi > \sim | (T-T_c)/T_c|^{.38\pm.01}
\ee
Recent analysis \cite{Karsch,DeTar}
has concluded, that the data are consistent with O(4) critical exponents,
 although say O(2) ones are not also excluded.

  The second obvious issue is the behaviour of global thermodynamical
quantities.
 The O(4) spin system has  an
amusing behaviour, with {\it positive} power
for specific heat\footnote{As far as I know, it
remains unknown whether the coefficient is positive or negative: thus one can
have a dip or a peak.}
\be C(T) \sim | (T-T_c)/T_c|^{.19\pm.06}
\ee
It means that the singular contribution of the soft modes
 {\it vanishes} at the critical point, and in order to single it out
 the 3-ed derivative
of the free energy should then be calculated.
Nevertheless,  lattice data for the  $N_f=2$ QCD
actually do show a {\it huge peak} in
the specific heat around $T_c$.  It certainly implies, that
many new degrees of freedom become available
(or are significantly changed) in this region.
What these degrees of freedom are, both in hadronic language and in the
quark-gluon one, remains the major open problem in the field.
 ( Of course, there is no logical contradiction here: apart of large but
smooth
peak one may eventually find a small 'kink', which is truly singular.

   Now we return to U(1) symmetry,
For simplicity, we consider only two light flavors and
use the old-fashioned notations, calling the isoscalar
I=0 scalar channel a $\sigma$ one,
and isovector $I=1$ scalar channel a $\delta$ one
\footnote{Now particle data table denote notations $f_0$ and $a_0$ to I=0,1
scalars: however particular resonances listed there under these names hardly
have anything to do with correlators under consideration.}.
Under $SU(2)_A$ transformations, $\sigma$ is mixed with $\pi$, thus restoration
of this symmetry at $T_c$ require identical correlators for these two channels.
Another chiral multiplet
is $\delta, \eta_{non-strange}$, where the last channel is the
SU(2) version of $\eta'$: at T=0 those are very heavy and are not considered
in chiral Lagrangians, or course.
$U(1)_A$ transformations mix e.g. $\pi,\delta$ type states,
and thus its 'practical restoration' should imply that  such type of
correlators should become similar. Finally, if {\it both} chiral symmetries are
restored, a simpler statement follows: left-handed quarks never become
right-handed, therefore all $\pi,\eta_{non-strange},\sigma,
\delta$ correlators should become the same.

 In their original paper Pisarski and Wilczek have actually argued\footnote{
They have even mentioned
 that this amplitude should  be at
$T_c$ at least an order of magnitude smaller than at
T=0, although no details of this estimate were
given.} in
 favor of the 'scenario 2'.
 Their argument was as follows:
'if instantons themselves are the primary chiral-symmetry-breaking mechanism,
then it is very difficult to imagine the unsuppressed $U(1)_A$-breaking
amplitude at $T_c$'.  However (as will show below) instantons do not seem to be
very strongly
suppressed at $T\sim T_c$.

  In ref.\cite{Shuryak_pt} I have argued that U(1) should be
``practically restored"
 right above
the transition region: the reason is instantons are forming molecules, rather
than being suppressed.
Let me now show that the latest lattice data indeed support this conclusion.
The argument will need some preliminary discussion.

    Suppose one is willing to measure masses of $\sigma$ and $\delta$
scalars on the lattice. If so, one should evaluate the so called ``connected"
and ``disconnected" quark diagrams. $\delta$ correlator has only connected one
( For its charged component e.g. $\bar u d$ it is trivial, because
two quarks have different flavor,
 a one-line algebra shows why it is so for neutral component as well.)
while disconnected diagram contributes to the $difference$ between them.

The available set of data \cite{LK} is shown
in Fig.6, as the so called ``susceptibilities",
the second derivatives of the free energy over quark masses.
In other words, it is the $integrated$ point-to-point scalar correlation
function: it is important, that the contribution of each scalae state is
therefore $inversely$ proportional to its mass squared.

 One can see that both
of them show strong T-dependence in the vicinity of $T_c$. The ``disconnected"
one, being a $ difference$ between $\sigma$ and $\delta$ correlators
first rises sharply (because sigma mass goes to zero at $T_c$) and then
rapidly drops. Unfortunately, there is no more data points, and therefore we do
not
know how small it actually become. If it is $really$ small, one may say that
$both$ chiral symmetry are practically restored, and (if it happens), it is
most
probably at the temperature only few MeV above $T_c$. We return to consequences
of this observation below, when we will discuss fluctuations in the critical
region.

\begin{figure}[t]
\vspace*{3.7in}
\caption{``Disconnected" and ``connected" scalar susceptibilities
(upper and lower
points) for 3 different quark masses, measured by Bielefeld group. The peaks,
marking the minimal $\sigma$ mass, correspond to $T_c$.  }
\end{figure}

\subsection{How to get experimental evidences for the phase transition?   }

   In this subsection we jump from theoretical considerations and numerical
experiments to a discussion of issues relevant for ``real" experiments.
All event generators, hydro and other models
agree  that at AGS (10-14 GeV*A) and SPS (200 GeV*A) energies
we should have reached the ``mixed phase" region (with large and small
baryon density, respectively). At the same time,
 ``pure" QGP is either not there, or
it exists for so small time that we (so far) cannot figure out how to
find its direct manifestations\footnote{This
point of view is not universally accepted. For example, H.Satz has repeatedly
suggested
 that the observed $J/\psi,\psi'$ suppression cannot happen in hadronic phase,
and thus some presence of QGP is needed to explain that.
J.Rafelski thinks the same is true about observed enhancement of multi-strange
baryons. I think both claims may well be
right, but more data (especially
with new lead beam) and more quantitative theory are needed
to accept them.
}. That is why  eventually RHIC and LHC heavy ion experiments will be
performed,
where QGP will be created way above the critical region and therefore be there
for significant time.

  Nevertheless, one should be able to find
some manifestations of the QCD phase transition in
current experiments, performed at Brookhaven AGS
 and CERN SPS.
A very direct  signature is
{\it transverse collective flow}, directly related with EOS.
 As noticed long ago \cite{SZ,transhydro}, near the QCD phase
transition the EOS is especially $soft$. As an illustration,
we show  Fig.7 . from \cite{HS}, where
 a $conventional$ parametrization of EOS is presented
  in an unconventional way. We have
 eliminated temperature
and plotted instead the hydrodynamically relevant ratio
$p(\epsilon)/\epsilon$ versus $\epsilon$,
thus emphasizing the existence of a
minimum at $\epsilon=\epsilon_{max}\approx 1.5\ \
{\rm GeV/fm^3}$. This minimum is referred to below as
{\it the softest point} of the EOS.

\begin{figure}[t]
\vspace*{3.5in}
\caption{Equation of state, plotted as the pressure-to-energy-density
ratio versus energy density.
The minimum is the ``softest point" of the QCD equation of state.  }
\end{figure}

  For long time only central collisions were discussed, with an axially
symmetric (cylindrical) flow. Generally speaking, the
AGS/SPS data agree well with soft EOS because
  no significant growth of the average transverse
momentum $<p_t>$ with the
 multiplicity (or transverse energy)
was observed. However, it is difficult to separate reliably thermal and
collective components of the particle momenta, and therefore really
quantitative analysis of flow in $central$ collisions is still missing.

Fortunately,
recent results from E814 \cite{814} have revealed flow
for $non-central$ collisions, as asymmetry  in the reaction plane,
similar to what was seen previously at BEVALAC.
One may expect to reach
quantitative understanding of the transverse expansion soon enough,
leading to restrictions for
the EOS.

\begin{figure}[t]
\vspace*{5.0in}
\caption{
Hydrodynamical evolution at two energies (from Hung and Shuryak)
on time t - longitudinal coordinate z plane.
Solid lines correspond to fixed temperatures, dotted ones to fixed longitudunal
velocity. M and H mark the domains of mixed and hadronic phase. Fig.(b)
corresponds to the long-lived fireball discussed in the text.  }
\end{figure}

\begin{figure}[t]
\vspace*{6.0in}
\caption{Energy dependence of
some observables, from Hung and Shuryak. Part (a) show lifetime of the mixed
phase (dotted line and right scale) and 4-volume of the mixed phase
(solid line and left scale). Part (b) show the hight (dotted curve, left scale)
and the width (solid line, left cale) of the dilepton rapidity distribution.
}
\end{figure}

   Let me now switch to the following general question:  which collision
energy is most suitable for observations of this ``softness" effect?
The answer considered in a recent work \cite{HS} seems quite obvious:
it is the collision energy at which matter is first produced close to
the
 ``softest point" indicated in Fig.7 above.
It was found that in this case one can see significant effect
not only the transverse, but also the $longitudinal$ expansion.
 Hydrodynamical studies of
central Au+Au collisions at varying energies were made, from the SPS to the
 AGS ones, 200 to 10 GeV/N.
Radical changes are found: from (i) violent longitudinal
expansion, close to scale-invariant solution at high energies, to (ii)
 a ``slow burning"
at the softest point; and then leading again to (iii)  a noticeable
 longitudinal expansion of  the hadronic gas
at low energies.
 In Fig.8,9 we have shown how the picture of the expansion and
some  global parameters of
 the space-time picture depend on collision energy.
The {\it maximal lifetime
 of the mixed phase}
(measured at $z = 0$)  $\tau_{mix}$ has a clear peak, corresponding to
initial conditions at which the pressure-to-energy ratio is the smallest.
The  total 4-volume of the mixed phase $V_M$ (also shown in Fig.9 )
has only a ``shoulder"; this is because longer lifetime is compensated by
smaller spatial volume.
Furthermore, these radical changes in the space-time evolution translates
into experimentally observable quantities.
The {\it penetrating probes}, $\gamma$ and $e^+e^-$ production are most
relevant here. Using
 production rates
\cite{photons} for photons and \cite{dileptons} for dileptons
(which can be compared to existing SPS data), it was found that the total yield
of  dileptons $dN_{e+e-}/dy(M,y=0)$ have non-monotonous
dependence on the  collision energy, with
 a very sharp rise near the
 ``softest point", with
more or less constant production level at higher energies.
Equally dramatic is the energy dependence of the $width$ (at half maximum)
of dilepton rapidity distribution.

 If the collision energy corresponding to the ``softest point" is
found, and the lifetime of the fireball is indeed factor 2-3 longer than
elsewhere,
one may suggest a whole list of interesting questions, related to
modifications of all
possible signals. To give few examples: Is there is any additional strangeness
enhancement, or $J/\psi$ suppression in this case?
\setcounter{footnote}{0}
     Other phenomena sensitive to the $total$ lifetime of the excited
system\footnote{
By the way, contrary to wide-spread opinion, the HBT correlations do not
measure it: they are only sensitive to the {\it emission} time of pions.
It is by no means the same thing: for example, fireball may exist for long
time, and then emit pions in a short flash.
}
are the so called low-$p_t$
enhancements. In my paper \cite{lowdens} it was suggested that the
collective potential may be able to $trap$ pions and kaons, provided their
transverse kinetic energy $m_t-m$ is smaller than the attractive potential.
(The phenomenon is similar, say, to  total light reflection when it comes out
of
water.)
The deviation from pure exponential (or pp collisions)
 was first seen for pions, leading to a series of different
explanations. Direct observations of multiple $\Delta$ resonances have
supported the resonance interpretation.
 In Quark Matter 1993 the results of
 E814 experiment \cite{814} has been reported: similar enhancement
was found
in spectra of negative and positive kaons, which
 have $no$ significant resonances near the threshold. What was
also surprizing, these first data, for SiAu collisions, have produced
the unusually small
inverse slope  $T\approx 15 MeV$ while more recent AuAu data
(see Fig.10 )
have shown much larger value. Why should this effect
 be {\it strongly projectile-dependent}?\footnote{
Note that
 at very small momenta one can see a dip in $K^+$ spectra, and also
a noticeable difference for $\pi^-,\pi^+$ ones:  it should be
related to collective Coulomb field: a good clock by itself.}.

\begin{figure}[t]
   \vspace{5.in}
   \caption{ Preliminary transverse energy spectra for
 $K^+$ masons and pions, reported by E877
experiment at QM95.}
\end{figure}

   Recall that for any amplitude, the energy derivative is related to
duration of the process under consideration: this is
basically  uncertainty relation. Thus,
   spectrum modification  at so small transverse energies as
$m_t-m$=10- 20 MeV
 implies  that the system lives long enough. A
particular
kinetic model  was used in \cite{Koch} in order to explain these
puzzling data. Indeed, it was found in the simulation that if
the fireball is expanded slowly enough,  the low-$p_t$ kaons
can be
$trapped$ in the fireball
and  ``cooled". In brief, the reason for cooling (the same in ordinary
refrigerator) is that kaons are reflected from the wall moving $outward$,
loosing energy.

  Let me present here a simple explanation, reproducing
at least these simulations
(if not data). No matter how deep is potential and how slow the motion is:
if it is slow enough one may
use {\it
 conservation of  adiabatic invariants}, which
accurately predict the behaviour of many similar phenomena, from
 atomic traps to expanding Universe.
 If a particle is trapped into an
expanding potential well
the following relation should hold
$ <p_t> R_t =const(time)$,
which for non-relativistic kaons translates into the following
``cooling relation"
\be
{T_{final} \over T_{initial}} = ({R_t^{initial} \over R_t^{final}})^2
\ee
The initial transverse
 size
$R_t^{initial}$ is that of the projectile nuclei, while the final one should be
that
measured by the HPT interferometry.
 For Si Au their ratio is known to be about 2.5,
so this formula indeed leads to nearly an order of magnitude drop in T.
For AuAu collisions the transverse expansion is less strong,
 leading to the final/initial size ratio of about 1.2, with much smaller
cxooling ratio.

 However,  many features
of this  picture remains unexplained. In particular,
large magnitude of the
observed enhancement can only be the case if the K rescattering rate
is miraculously reduced, so that they
may remain `cool' inside the hot fireball! Furthermore,
it was found in \cite{Koch} that in order to get that one needs
 fireball lifetimes at least 30-40 fm/c. It is several time more than
event generators suggest, and only  comparable
to what we get above for the long-lived scenario at the
``softest point".  Does it mean that such scenario may actually
take place at AGS?

  Another set of so far unexplained data is related with ``penetrating probes",
photons and especially dileptons.  Those
observables clearly
is the best ways to get information about the
early hot stage
of nuclear collisions  \cite{Shuryak_78}.
 Unfortunately, such experiments are difficult,
 so
only recently their first preliminary results
were reported by 4 CERN experiments:
 WA80 (photons),  NA34/3 and NA38 (dimuons) and CERES
(dielectrons). (See \cite{Tserruya} for recent review.)
 All of them see significant excess over the expected background
effects due to hadronic decays: in the CERES\cite{CERES}
 case at dilepton masses
$M\approx .3-.5 GeV$ it reaches one order of magnitude, as it is shown in
Fig.11.
\begin{figure}[t]
   \vspace{2.5in}
   \caption{ CERES preliminary data on dielectron mass spectrum, reported atr
QM95. Curves correspond to ``cocktail" of known decays, they explain p-Au data
but not the S-Au ones.}
\end{figure}

  Attempts to explain these observations using conventional
 dilepton production mechanisms (such as the fundamental
$\bar q q$ annihilation in the QGP phase,
    $\pi^+\pi^-$ annihilation in the hadronic and mixed phase, as well as
other important reactions involving
the $A_1$ meson) has been made by many groups, but they do not actually
 suceeded to get quantitative explanation
of the effect. In ref.\cite{SX_where} we have suggested
that again the reason might be much longer lifetime of the fireball, 30-40 fm/c
instead of conventional 10. Unfortunately, after hydro calculations \cite{HS}
were completed, it turns out that even the long-lived fireball produced at
the ``softest point" does not really help here, bacause larger lifetime is
compensated by its smaller spatial volume.

\subsection{Hadron modifications at finite temperatures}

   It is very natural to expect the elementary excitations (hadrons)
to change their properties at non-zero temperatures/densities, especially close
to the phase transition. We have discussed above lattice data on some scalar
masons, $\sigma,\delta$: do we have any experimental evidences from `real"
experiments as well?

   This question was studies for long time in normal nuclear matter, for
nucleons. In short, large amplitude scalar and vector collective fields are
found, which however nearly cancel each other in the nucleon case ( and we do
not understand why). What about other hadrons?

  One of the most interesting case is
that of $K$-mesons (we have discussed it just above). Recent analysis
of K$^-$ atoms \cite{Gal} have demonstrated
very large attractive potential about $-200$
MeV inside the heavy nuclei. If correct, it makes  kaon
condensation in stars \cite{Kcond} unavoidable, and kaon ``trapping" inside the
fireball of expanding matter very easy.
There are indirect evidences of $\rho$ modification as well: and this is about
all.

   There are several different ways how one can  approach the problem
of hadron modification theoretically, such as:
(i) chiral perturbation theory;
(ii) direct lattice calculations; (iii) rescattering corrections based on
phenomenologically known scattering amplitudes; (iv) QCD sum rules;
(v) effective Lagrangians; (vi) other models, including ``instanton liquid";
etc. Let us discuss them subsequently.

{\bf Chiral perturbation theory} can generally be used
to describe  interactions of soft pions: and at $low$ T
this is the case. Classic examples include Gerber and Leutwyller
 expression for the
quark condensate \cite{Leutwyller}
\be
{<\bar q q (T)> \over <\bar q q (0)>}=1-{T^2 \over 8 f^2_\pi}
-{T^4 \over 384 f^4_\pi} +...
\ee
or the general theorem \cite{Eletsky1} according to which $O(T^2)$
corrections to hadronic masses are absent.
(In the next $O(T^4)$ order such shifts however appears
\cite{Eletsky2}.)
\setcounter{footnote}{0}

 New result, obtained in similar way, is the low-T correction to the
instanton density, recently evaluated in \cite{SV_lowT}. We will discuss it
below.

{\bf Lattice calculations}
are covered in recent review \cite{DeTar,Boyd}.
We have partly discussed those above, in connection with the softest
modes, the pseudoscalars and scalars: those particles
do indeed show large mass
modification.

 The most interesting channels from experimental point of view are vector ones,
especially $\omega,\phi$, because their modification can be measured with high
accuracy and also because their lifetime is comparable to that of the fireball,
so they are, so to say, natural clocks.
 There are some preliminary studies of
$vector$ mesons, so far in the quenched approximation: from those one may
conclude that they are $not$ shifted noticeably till $T\approx .9 T_c$.
However, it remains unknown what happen with vector mesons in the ``mixed
phase", where hadronic system produced in nuclear collisions spends most of its
time.

   Lattice people has invented another way of looking at hadronic modes,
measuring hadronic correlation functions in $spatial$ rather than $temporal$
direction. They have some poles, known as ``screening masses". Those exist
also well above $T_c$, and should not be confused with
 modification of
   hadrons
discussed above\footnote{Propagation in time and
space directions are related to $electric$ and $magnetic$
forces, which are affected by QGP in a different way.
Electric ones are strongly screened in the lowest order, while magnetic ones
are probably screened non-perturbatively. Also confinement effects work
differently: for paths going in space directions there is no ``deconfinement"
till any T.}.
 In Fig.12 we show a sample of lattice measurements
compiled by Gocksch\cite{Gocksch}: at large T one can see that
these masses tend to $2\pi T$ for mesons and $3\pi T$ for baryons.
We will return to discussion of the screening masses below.

\begin{figure}[t]
   \vspace{4.in}
   \caption{ Compillation of lattice results on ``screening masses" versus
temperature, from Gocksch. }
\end{figure}
\begin{figure}[t]
   \vspace{2.5in}
   \caption{ Real and imaginary optical potential for $\omega$ meson moving in
the pion gas,
versus its momentum p.  Three curves correspond to T=150,175,200 MeV.
Curves for other mesons look similar. }
\end{figure}

{\bf Rescattering corrections based on
phenomenologically known scattering amplitudes:}
such way of calculation is quite traditional. For hadronic
 gas it was explored in
papers \cite{lowdens}, where for $\pi\pi$ and many other channels
the well known phase shifts
can be explored. What is important, in this way
one gets not just the mass or width shift, but actually modification of the
whole dispersion curve $\omega(k)$ for excitations. It may be quite
non-trivial: for example, pions (and other pseudoscalar mesons)
 are protected {\it at small} momenta by
the Goldstone theorem: they do not interact much. However, at larger momenta it
is
no longer so, and one may expect much larger modifications.

In general, such approach leads to relatively small ``collective potentials",
of the order of nuclear potential in nuclear matter.
As a sample of the results, let me show the calculated optical potential
(both real and imaginary parts) in Fig.13  for
$omega$ meson, versus its momentum p. This modification should be directly
observable in the dilepton channel at RHIC by PHENIX detector, which has
sufficient mass resolution.

{\bf QCD sum rules at finite temperatures}
Basically, this approach relates hadronic masses with the quark condensate:
and as it is dropping near $T_c$, one expects hadronic masses to decrease as
well. In generally, it seems to be very reasonable idea. However, in practice
its implementations has met some problems.

   First of all, the original papers \cite{BS} have confused technical issues
(see discussion in \cite{Shuryak_cor}).
A fictitious ``T-dependent  OPE coefficients"
 were introduced (even for  unit
operators ), while many relevant operators (those which are
not Lorentz scalars) were omitted. Later on, those points were corrected, see
e.g.
\cite{Hatsuda}.

 Another general problem, undermining predictive power of QCD sum rules
is very small (and not quite understood) region of validity.
OPE provide the correlators at small distances only, to start with, and
it is well known that in certain channels it does not work at all. Those are
scalar and pseudoscalar channels, the nucleon one and others where ``direct
instantons" can contribute.
\setcounter{footnote}{0}

  By the phase transition point $T\approx 150 MeV$ the $\rho$ meson mass
was found to be shifted by about 10\% $down$, while the $\omega$ mass is
predicted to remain the same. (This contrasts my estimates based on
scattering amplitudes, which found similar shifts for both $\rho,\omega$:
future dilepton experiments hopefully can tell the difference.)
Asakawa and Ko \cite{AK} has calculated the shift of $\phi$ mass, and again by
the phase transition point they have predicted it to become
 reduced by about 10\%. If so, experimentalist clearly have possibility to fond
a ``second peak" in dilepton spectra.

Let me also mention that in some cases one can prove exact relations,
which should be obeyed by spectral densities at $any$ T. Those are for examples
analogs of Weinberg sum rules, for vector minus axial correlators, derived in
 \cite{QCDsumrulesT}.

{\bf Effective Lagrangians} are well known tools, and respectively there are
many papers on the subject. The simplest proposal is Brown-Rho scaling,
according to which for $all$ particles
\cite{Brown_Rho}
\be
{m(T)\over m(0)}  = [{<\bar q q (T)> \over <\bar q q (0)>}]^{1/3}
\ee

Next, there are extensions of sigma-model. Recent paper by Pisarski
\cite{Pisarski} is based on gauged version of it, so one can calculated
shifts of $\rho$ and $A_1$. His predictions for $M_\rho(T)$
are as follows: it first starts decreasing (in the chiral limit,
proportional to $T^4$, of
course), but then grow again, reaching
at $T_c$ $\sim
962 \, MeV$.  However, $M_{A_1}$ does exactly the opposite,
it grows and then decreases (of course,
 meeting the rho mass at $T_c$).
The width of the thermal $\rho-a_1$ peak is estimated to be about
$200 - 250 \, MeV$ at this point. So, at low T the behaviour
 resembles simple repulsion of two levels which are being mixed:
 from that perspective one may think that effect of all
unaccounted states (e.g. of $\rho'(1600)$ on $A_1$) will be to push them
downward.

  All
 applications of effective theories have one basic problem: by using them
one specifically $assumes$ that
 all coupling constants (as well as ultraviolet cutoffs
etc) to be T-independent.
(Otherwise, there predictive power is lost.)
 But why should  it be the case?

{\bf Microscopic models }
try to explain how hadrons are created, and thus can provide some insight
into T-dependence of such parameters. For example,
the ``instanton liquid" model does generate NJL-type interactions,
and we know that those have strength proportional to the instanton $density$,
and cutoffs related to instantons $sizes$. As we will discuss below,
we have some information about their T-dependence.

\subsection{Critical fluctuations}

   Another general idea, discussed in many papers, is to try to find
large-amplitude fluctuations in data,  similar to critical fluctuations known
in
many other physical systems. Let me skip discussion of fluctuation in
hadronic reactions (including the pp ones) as too vast subject, and only
comment on specific proposal related to the QCD phase transition.

    Evaluation of ``nucleation rate" of hadronic bubbles in $supercooled$
 plasma was discussed by Czernai and Kapusta \cite{bubble1}. Their main
assumption is that the transition is of the first order and the main result is
that supercooling cannot be too strong, so that the system returns back to
the mixed phase as determined by Maxwell construction. At the same time,
supercooling $\Delta T$ seems to be large enough, more than 5 MeV or so
in which lattice data seem to be safely confined: it means that this work would
not be affected if the transition happen to be actually a rapid
crossover. The major problem of this work (discussed in fact by its authors
themselves) is a relative $smallness$ of nucleation rate compared to realistic
lifetime of the system.

   Related discussion of nucleation, this time for $overheated$ hadronic gas
going into the plasma phase, was done in \cite{bubble2}. Again, classic
thermal excitation
formulae give too small rate, suggesting
probability for central AuAu AGS collisions to create QGP to be
 of the order of 1\%.
However,
   in real liquid-to-gas transitions we know that some
small
inhomogeneous perturbations (rough surface,ions or other dirt, etc)
 actually dominate nucleation,
and we have to find their analog in heavy ion collisions.
 Work on such ``seeds"
is in progress\footnote{Vischer and Kapusta, private communication.},
 and that now they hope to get much larger probability to
produce QGP.

  Still it is natural to imagine, that if we are somewhere in the mixed phase
(which is certainly the case at AGS), not 100\% of the events follow
the same trajectory on the phase diagram. How can we separate those?
I think a possibility is to think about
``softness" of EOS and related flow (which is determined
for non-central collisions on event-per-even basis, more or less). If
the event with overheated mesonic gas is found, which suppose to have
 very different and  $hard$ EOS, therefore the flow might be much stronger.

  The  idea of {\it disoriented chiral condensate} (DCC)
 \cite{DCC} has created much theoretical work\footnote{And even
the FNAL dedicated experiment, lead by Bjorken himself.}.
In short, it suggests that the
 newly formed bubble of hadronic matter imbedded in QGP
does not know the ``politically correct" direction of the quark condensate
in the isospin space (that is prescribe by relatively small quark masses)
and can have random orientation instead. A consequence is large fluctuations in
neutral-to-charged pion ratio, especially in the bin of small $p_t$.

The major proposal was put forward by Rajagopal and
Wilczek \cite{RW} who has shown that if
 cooling is very rapid (``quench") one has instabilities
of lowest modes, which may grow significantly.
 (That was indeed observed in numerical studies mentioned.)
However, it remains completely unclear whether in fact such a quenched
scenario can be the case in heavy ion collisions.
In fact, very large jump in entropy and energy density at $T_c$ \footnote{
In fact, it is not due to pion and sigma fields, of course, and was not
included in simulations based on sigma model.}
 force the system to stay very
$long$ time near $T_c$.
For example, the duration of the ``mixed phase" is expected to be 20-30 fm/c
at RHIC, to be compared to instability increment $1/m_\sigma \sim 0.3 fm$.

  My own ideas about DCC are related with the ``practical" U(1) restoration
discussed above.
  Large fluctuations in $\pi,\sigma$ directions should exist right above $T_c$:
but the same should be true for
their U(1) partners, $\delta,\eta'$.  Unfortunately, unlike vectors
those excitations do not decay into something which is
possible to see directly. However, if they are $trapped$ inside the DCC bubble
by their large masses outside, they may
for example significantly enhance the bubble lifetime.
Much more work is needed in order to figure out whether we have chances to see
this phenomena.

\subsection{Instantons at finite temperatures.}

   We have claimed in chapter 2 that instantons dominate hadronic physics
in the vacuum. If so, they should also produce important effects at non-zero
temperatures, below and even above $T_c$. Unfortunately, this subject was not
yet studied in sufficient details, and only first steps have been made, to be
reviewed in this subsection. But before we do so, some technical points are
needed.

   First of all, a finite temperature is introduced in a remarkably
simple way (invented by Matsubara long ago) in Euclidean formulation
of quantum field theories: just  ``time" direction simply
become finite, with the length 1/T.
 Second, it is quite straightforward to generalize the instanton
solution of Yang-Mills equations to this case: one should simply look for a
periodic solution, or a periodic set of instantons.
We would not show the corresponding formulae here, and only comment
about the limiting cases. Naturally, at small temperatures, when the box size
is very large, it leads to relatively small deformation of instantons, compared
to its original 4-dimensionally symmetric form at T=0.
At high T, on the contrary, all traces of space-time symmetry are gone.
At small spacial distances the solution develops a universal
(=$\rho$-independent) ``dion" field, with static electric and magnetic fields.

  It is useful to remind here, that in the rest of this
chapter we will consider mostly the region close to the QCD phase transition.
Therefore,
the relevant box size is supposed to be about $1/T_c\approx 1.3 fm$, roughly 4
times the mean instanton radius, $\rho=1/3 fm $.
So, the instantons themselves can be well fitted into the box, without too much
deformation.

   How the {\it instanton
density} depends on the temperature? Clearly,
one should know that in order to understand the role they play in his case.
At $large$ T  it is well known \cite{Shuryak_conf,PY}
 that  only the small-size
instantons such that $\rho  < 1/T$ can survive, because essentially of
the Debye screening of their field.
  It was argued in ref.\cite{SV_lowT} that such suppression of instantons
should only exist
above $T_c$, because it is essentially a Debye
screening.

At
 $small$ T
was worked out only recently, in \cite{SV_lowT} using the PCAC methods.
The result (for two massless flavors)
is expressed via {\it vacuum} expectation values of two
complicated 4-fermion operators, which have $calculable$ T dependence,
namely $[1- {T^2 \over 6 F_\pi^2}]$ and $[1+ {T^2 \over 6 F_\pi^2}]$
(our pion decay constant is $F_\pi=93 MeV$).
To start with, this dependence is weaker than for the condensate squared
($1- {T^2 \over 4 F_\pi^2}$), and
available estimates of the vacuum average values even suggest
that it may even cancel out, when the two terms are
added.

\setcounter{footnote}{0}

  The relevant lattice data
are very very crude so far,
the best \cite{Chu_Schramm} are shown in Fig.14. One can see that
both conclusions seem to be justified, in particular suppression
seems to agree with Pisarski-Yaffe formula, provided one substitutes
$T^2\rightarrow (T^2-T^2_c)$. Also, in ref.\cite{Chu_Schramm} there are
evidences that instanton size is constant below $T_c$, but it decreases above
it.
\begin{figure}[t]
\vspace*{3.5in}
\caption{Density of instantons as measured by the ``cooling" method versus
temperature T, from Chu and Schramm (see text). PCAC stands for Shuryak and
Velkovsky low-T limit, and P-Y for Pisarski-Yaffe high T one. }
\end{figure}

   Now, how the instanton {\it interaction}
  changed with
temperatures? This is the topic discussed in details  in \cite{SV_temp}.
The most drastic changes happen to be with quark propagation. At high T, the
corresponding zero modes look approximately as
\be \psi_0(\tau,r)=
sin(\pi T \tau)/cosh(\pi T r) \ee
where $\tau,r$ are distance from the center in time and space direction.
Note a crucial difference between the  dependence on time and space distance:
oscillations in time versus exponential decay in space. The latter
are due to the famous fact: the lowest Matsubara frequency $\pi T$ for fermions
are non-zero, due to anti-periodic boundary conditions.

  In our discussion above, we have compared ensemble of instantons with some
``liquid" made of atoms, with quarks playing a role similar to electrons.
Using this language further, one may say that our ``atoms" becomes
more and more anisotropic, as the temperature grows\footnote{ That is similar
to what
happens with ordinary atoms in very strong magnetic field (e.g., on a pulsars),
in which the Larmore radius is smaller than the Bohr one.}.
As we will see below, such deformation will radically change properties of
their ensemble.

 The main phenomenon in this region is
 a strong ``pairing" of instantons, leading to
splitting of the instanton liquid into a set of $\bar I I$ molecules.
The first (strongly simplified) discussion of chiral restoration
transition
at this angle was made in \cite{IS}.
\setcounter{footnote}{0}

   New finding is strong and rapid ``polarization" of these molecules
in the critical region. We have already discussed the main anisotropy
of the interaction, which
comes from the quark-induced interaction. Consider an instanton at the origin,
and an anti-instanton with a center placed at distance $\tau,r$ is time and
space directions. If they form an isolated system (a ``molecule") and we are
discussing the theory with $N_f$ type of massless quarks\footnote{
The QCD can be considered as a case between  $N_f=2$ and $N_f=3$,
something like 2.5.}
 then the fermionic
determinant should be proportional to
\be
  det D \sim |sin(\pi T \tau)/cosh(\pi T r)|^{2N_f} \ee
because a pair of each quarks should travel from one to another.
(We have used here an approximate form of the zero mode considered above.)

 Note that the point $r=0,\tau=1/(2T)$ is  a strongly peaked
 maximum of this function.
It corresponds to ``polarization" of the molecule in time direction,
and a particular position in time direction, for which both centers are at
 the opposite sides of the torus.

  In order to see how important is this configuration for a {\it single}
molecule, we made a simulation (of course, with realistic masses and
more accurate expressions). The results
 demonstrate that the
 degree of polarization rapidly grows in the vicinity of T=150 MeV,
or exactly at the point of chiral phase transition.

    Roughly speaking, the critical temperature is then defined as the size of
the Matsubara box, such that an ``molecule" can be nicely fitted in, in the
 time direction.
In more strict sense,
one can say that at the transitory region $T\sim T_c$
the ``instanton liquid" is changed into a kind of ``liquid
crystal" of nematic type. Instantons are being ``married" into closed pairs,
 therefore they
stop communicated with each other, which in turn leads to
  the disappearance of their
common ground, the quark condensate.

  Let me add a comment on the physical meaning of the
polarization phenomenon in a less technical language. The $I\bar I$
molecules are a virtual (or failed) tunneling event, in which the
gauge fields penetrate into a new classical vacuum only for a short period
of time, and then return back. For that reason, they do not contribute
to the quark condensate and other related quantities. ``Polarization" of the
$\bar I I$ molecules at $T\sim T_c$  means that
at such temperatures the tunneling is concentrated in
the vicinity of {\it the same spatial point}.

  Before we consider this problem more quantitatively,
 let us  make one more digression.
Even if the contributions of ``molecules" in the QCD vacuum
and the phase transition is not as large as I think it is,
there is an external parameter that  can increase their role.
The way of doing this is to
increase the number of light flavors $N_f$. As the fermionic determinant
is raised to a higher power $N_f$, the role of correlations induced
by the determinant certainly increases.
 Thus, one may anticipate larger role of molecule-type correlations
even at zero T, and smaller $T_c$ in this case. Lattice data do indeed
suggest decrease of $T_c$ for larger $N_f$. Furthermore, there are
(so far not very convincing) data about
existence of some critical number of flavors $N_{f,\,cr}$ above which
chiral symmetry breaking and confinement are absent even in the ground state.
Whether it is indeed so, and whether this new phase of QCD-like theories can be
described by
the instanton ensemble dominated by molecules, still remains to be seen.

\subsection{Instanton-induced interactions and the equation of state}

   The results discussed in the preceeding section has
significant impact on physics around the phase transition point $T_c$
 and above it: if instantons do not disappear there, they generate
non-prturbative forces.
Furthermore, ``molecules" generate forces quite different from those for random
ensemble.

   The Lagrangian for these new interactions was considered
in \cite{SSV_mix}. To model the ``mixed phase", a schematic
``cocktail model"
  was used,
containing both random component and some
fraction of molecules  $f=2 N_{molecules}/ N_{all}$.
We have first found that $<\bar q q>$,
depends  on $f$ in a way similar to
its T-dependence, measured on the lattice: it changes little first, and then
rapidly vanishes at $f\rightarrow 1$.
Different correlation functions depend
on $f$ quite differently. For example,
  for $\pi,\rho$
channels one finds
remarkable stability for $f=0-0.8$, with subsequent strong drop
toward $f=1$. At the last point {\it complete}
 chiral symmetry gets restored, so the pion correlator
 coincides with
its scalar partners $\sigma,\delta$.

  In Fig.12 we show a sample of results for ``screening" masses
\cite{SSV_mix}, for pion, rho,
nucleon, $a_1$ axial meson, and $Delta$. One may see that
although the temperature was kept to be $T=T_c=150 MeV$, for all channels
 except pion one the high-T limits
 $2\pi T$ and $3\pi T$, corresponding to lowest Matsubara
frequencies, are actually reached. Similar calculations for interacting
ensemble (where the fraction f is detemined by the statistical sum itself),
and for ``real" masses are now in progress.

\begin{figure}[t]
\vspace*{4.5in}
\caption{ ``Screening masses" for correlators with pion,rho, $a_1$, nucleon
and delta quantum numbers in the instanton ensemble, possessing fraction
f of ``paired" instantons. Note that at f=1 chiral symmetry is completely
restored (e.g. $\rho,a_1$ masses are the same).
}
\end{figure}

   Finally, let us return to a very difficult problem (mentioned already at the
very beginning of the Introduction), that of non-perturbative vacuum energy
density.
 Recall that, in terms of (Minkowski) field strengths, it is
\be \epsilon={1\over 2}(E^2+B^2)+ g^2{(11/3)N_c-(2/3)N_f \over 128\pi^2}
(E^2-B^2)
\ee
The first (Maxwellian) term is classical, and the second is ``anomalous" due to
quantum corrections.

  This expression can be compared to
 a bag-type expression
$
\epsilon_{vac}= \epsilon_{perturbative} - B
$.
Perturbative energy density is related with both classical and anomalous terms,
it
 is badly divergent but those infinite
parts are just additive constant which can be subtracted.
(In other words, we put ``perturbative vacuum energy" to be zero
 by convention.)
Instantons
have imaginary E and real H which cancel each other in  the classical
term,
but the second  one contributes.
This is the ``vacuum energy" shift due to tunneling we mentioned in the
Introduction.

  At high $T>>T_c$ the perturbative part gets its Stephan-Boltzmann
$\epsilon,p\sim T^4$
contribution of the QGP (times the perturbative corrections).
The nonperturbative phenomena also become T-dependent. In particular,
for the
 ``polarized" molecules one finds $E^2=-.8 B^2$. Thus, the first
classical term works, producing $positive$ energy, which is actually
large, of the order of 1 $ GeV/fm^3$!

 The issue is what happens in the vicinity
of $T_c$. On general grounds one can see, that certain scenarios
are impossible. For example, if all instantons would disappear instantly,
$at$ $T=T_c$, then thermal pressure of QGP should be sufficiently large
to compensate that loss, because {\it p(T) cannot be a decreasing function},
$s(T)=dp/dT>0$.
Since for QCD
with dynamical quarks the
 critical temperature is rather low, $T_c\approx 150 MeV$, this condition
is $not$ actually fulfilled. Thus, the B term $cannot$ instantly disappear!

  As it was discussed in the preceeding sections, instantons do not disappear
indeed, and generate important effects even above $T_c$. How exactly it
happens remains unknown.
   In reference \cite{IS2} a schematic model was developed, which provide some
example, and also show
  whether the bag-type model makes any sense.
 In this model the instanton
ensemble is described as a mixture of a molecular and a random
component. The partition function for the two components
are assumed to be
\be
 dZ_m&=& C^2d\rho_1d\rho_2d^4RdU\,
      (\rho_1\rho_2)^{b-5}\exp\left[ -\kappa (\rho_1^2+\rho_2)^2
      (\overline{\rho_a^2}n_a + 2\overline{\rho_m^2}n_m)\right]
      \, \langle\left( T_{I\bar I}T^*_{\bar II}\right)^{N_f}\rangle
\ee
for the molecular component and
\be
 dZ_a&=& 2C d\rho\, \rho^{b-5}
       \exp\left[ -\kappa \rho^2
      (\overline{\rho_a^2}n_a + 2\overline{\rho_m^2}n_m)\right]
      \, \langle TT^\dagger \rangle^{N_f}
\ee
for the random component. Here, $n_a,n_m$ denote the densities
of the random and the molecular components, $\overline{\rho_a^2},
\overline{\rho_m^2}$ are the average square radii of instantons
n the two components, $C$ is the normalization of the single
instanton density, and $b=\frac{11}{3}N_c
-\frac{2}{3}N_f$ is the coefficient of the Gell-Mann-Low function.

   The model uses a simplified gluonic interaction corresponding
to an average repulsion $\langle S_{int}\rangle= \kappa \rho_1^2
\rho_2^2$ parameterized in terms of a single dimensionless constant
$\kappa$. The fermion determinant for the random component is
approximated by the average $\angle TT^\dagger\rangle $ of the overlap
matrix element $T_{I\bar I}$ averaged over all positions and
orientations. For the molecular component, on the other hand, the
overlap matrix element is first raised to the $N_f$ power and
then it is averaged over all  positions, whereas the relative orientation
is kept fixed.

   Thus, as one can see, the $only$ element of the model depending on the
temperature T is the
quark-induced interaction. Remarkably enough, it is sufficient to
 generate the
chiral phase transition, at about the right T. It happens as follows:
 the average value for the quark determinant
 gradually {\it decreases} with temperature for the random component, whereas
the
determinant for the molecular component {\it first increases}
(at $T\sim T_c$) and
eventually, at larger T, starts to {\it decrease}.

   In Fig.16 we show a sample of results\cite{IS2}
 for the resulting thermodynamical
quantities, and one can see that a significant portion of the jumps at $T_c$
is due to instanton contributions.

\begin{figure}[t]
\vspace*{5.in}
\caption{ Instanton density, pressure and energy density versus temperature T
in the schematic model described in the texts. Left and right panels are two
variants of the model, showing uncertainties invilved. For n panel, solid line
is random componentm and the dashed line is the density of molecules. For p,
solid line is total pressure, decomposed into quark-gluon one (dotted lines)
and instanton contribution (dash-dotted one). For energy density (bottom),
solid lines are total sums, and dash-dotted ones show the instanton
contribution.
}
\end{figure}

\section{Formation and equilibration of quark-gluon plasma}

\subsection{Main predictions for  RHIC energies}

Already  the first papers where search and signals for quark-gluon plasma
were suggested  \cite{Shuryak_78} has actually  addressed the issue
of parton thermalization at high energies\footnote{Let me remind that
at that time it was assumed that
 CERN ISR could become a major facility for high energy heavy ion studies:
unfortunately those were later cut off by the decision
 of CERN leaders to destroy
it.}, and  the main qualitative features of what later
became known as ``the hot gluon scenario"  were
proposed.

 In particular, it was already recognised that
going from low (AGS/SPS) energies to RHIC ones one enter new domain, in which
 soft hadronic physics play smaller role, while
processes, which involve partons with momenta
 $p\sim$ 1-3 GeV (known also as
'mini-jets') are in fact dominant. Later those were studied in details
\cite{Kajantie_minijets1,Kajantie_minijets2,Eskola_etal,Blaizot_Mueller,Wang_92}
for pp collisions, and extrapolations to nuclear collisions were attempted.
 An additional complication compared to the pp
case is that in heavy ion collisions they
can no longer be considered as
isolated rare events, but a part of a  complicated ``parton cascades".

  To make a benchmark, let us recall what was called a
{\it standard scenario}\footnote{It was considered to be standard for about a
decade, in 80's.
}, which is simply based on
Bjorken's guess about the
{\bf equilibration time}   $\tau_0=1$ fm.
As one knows from pp,pA  data
the rapidity density of secondaries, one can extrapolate
to nuclear collisions. Another guess is multiplicity extrapolation:
 we use for central AA collisions
\be {dN_{AA} \over dy} = A^{\alpha}\,
0.8 \, lnE_{cm} \ee with
$\alpha=1.1$. The
entropy conservation leads then to
   the following initial entropy density:
\be s_i={3.6 dN/dy \over \pi R_A^2 \tau_0} \ee
(where
3.6 comes from the entropy/number density ratio for the {\it pion gas}
at breakup) and conclude that in this scenario
for {\it central} collisions at
RHIC (Au Au $\sqrt s=200$ GeV*A) and LHC (PbPb $\sqrt s=6300$ GeV*A)
the initial temperatures $T_i \approx 240;290\, MeV$ at RHIC and LHC.

   Moving forward,
    let us try to $estimate$ the kinetic equilibration time, using
   partonic kinetics.
The relevant  cross sections in the lowest QCD order
 are known to be
\be {d\sigma\over dt}={\pi\alpha_s^2 \over s^2}M^2 \ee
\be M^2_{gg\rightarrow gg}={9 \over 2}(3-{ut \over s^2} -{us \over t^2}
-{st \over u^2}), \,\,\,
M^2_{gg\rightarrow \bar q q}={1 \over 6}{(u^2+t^2)^2 \over u^2 t^2} -
{3 \over 8} {u^2 + t^2 \over s^2} \ee
\be M^2_{qg\rightarrow qg}=-{4 \over 9}{u^2+s^2 \over us} +
 {u^2+s^2 \over t^2}, \,\,\,\,
 M^2_{q_1q_2\rightarrow q_1q_2}={4 \over 9}{s^2+u^2 \over t^2 } \ee
where the subscripts in the last formula mean that two quarks are of different
kind, so the cross diagram is absent. The last expression
  also holds for $q_1 \bar
q_2$ scattering.

$Large$ angle cross sections\footnote{Small angle ones are larger, but they
contribute less to momentum equilibration.}
are very different: at $90^0$ the
 $M^2$ for these 4 processes are related as
\be gg/gg\rightarrow \bar q q/qg/qq= 30.4/0.14/5.4/2.2 \ee
 so  the gg scattering
\footnote{
In the gg
case an extra
factor 1/2 can be used, reflecting twice smaller t range:
 one should not take into account
the same final state twice.}.
{\it significantly exceeds} other processes, especially quark production.
Thus
one should expect a {\it two-stage equilibration},first of gluons with
noticeably {\it higher } $T_i$, and
later of quarks, with  {\it smaller} $ T_i$.

The "equilibration time" can be defined in many ways, let it
be the  time during which
{\it each parton has been in average
scattered  once}\footnote{Note that it is essentially the same condition
as  traditionally used for defining final (or breakup) parameters:
 the  system size is comparable to constituent
 mean free path.}. That leads to the following ``selfconsistency
 equation"  \cite{Shuryak_twostage}
\be  \tau_0= { 3.6 dN/dy \over \pi R_A^2}{1 \over 7.0T_i^3} \approx
\tau_g = {1 \over const T_i} \ee
(The  factor 7.0 comes from the entropy of the gluonic plasma
at $T=T_i$, and the constant in the r.h.s.
 should be taken from the scattering rates discussed above.) $Assuming$ the
same total multiplicity as above,
 one gets the initial gluonic temperatures
and the equilibration times
$$ T_g \approx 500\, MeV \, \, \, \tau_g\approx0.3 fm \, \, \, (RHIC) $$
$$ T_g \approx 660\, MeV \, \, \, \tau_g\approx0.25 fm \, \, \, (LHC) $$
(Here the effect
of  small-angle scattering is also included.) Note that these predictions are
significantly different from the ``standard scenario" mentioned above.

Observable consequences of this ``hot glue"
scenario include charm production. It was proposed
  as signature for high-T
 QGP  in \cite{Shuryak_78}. The mechanism is
 $gg \rightarrow \bar c c$ reaction, and
 its implementations for nuclear collisions
  were studied later in great details.
Direct (parton model) charm
production
 results in about 1 $\bar c c$/event (RHIC), while
 "thermal" production  leads to
$\sim 10^{-2},1,10.\bar c c$/
event at $T_i=300,400,550 MeV$.
Therefore, one can expect {\it significant increase of charm
production}\footnote{Considerable confusion has been created in literature
in relation with the so called ``intrinsic charm excitation". In
particular, K.Geiger \cite{PCM} has predicted it to be the dominant mechanism
of charm production, dominating by a significant factor over the thermal
production. However, I think those calculations strongly overestimate the yield
because most of the gluons are not virtual enough to resolve charmed pairs.
Further work is needed to get quantitative results.   }
compared to the scaled pp estimates.

 Spectra of
photons and dileptons produced in this scenario should also be significantly
different from those in the "standard" one:
during the "transitory time period" ($\tau_g < \tau < \tau_q$)\\
 one has {\it smaller} number
of quarks, but those are {\it hotter}: the $gg\rightarrow\bar qq$ process
is mainly active at {\it
 small angles}, so quarks simply have
  the temperature of gluons,
 As most photons and dileptons originate from the
tails of the distribution functions, it is important that their
relaxation to the equilibrium ones happens {\it from above}.

   Our discussion above $assumed$ the amount of entropy: and the treatment was
made so to say backward in time. Of course, one should be able to $evaluate$
how much entropy is produced directly, considering parton kinetics.
   Although qualitatively all approaches agree with
the ``hot glue scenario" outlined,
the detailed predictions for the $entropy$ produced are very different.

  One approach, based on  $binary$ scattering processes at the first impact,
is the HIJING model \cite{Wang_92}, which was formulated as an event generator
and therefore widely used. It was  supplemented by consideration of
radiative ``energy losses" in medium, and in this sense contain
some multi-gluon interaction
in the small-angle (leading log) approximation.
 However, rescatterings and  multi-parton processes
are not yet included in this model.

   The quantitative treatment of subsequent rescattering was attempted in
the ``parton cascade model"
(PCM) by Geiger and Mueller \cite{PCM}, which
aims to trace the partonic system evolution
all the way, from the
structure functions of colliding hadrons to final hadronization
of emerging mini-jets.
This model hives the highest numbers
for produced entropy,
because this model includes collisions with very soft gluons,
$x\sim 0.001$, which also happens $prior$ to real first impact.

It is based on {\it sequential branching
  of  virtual } partons (e.g.
 $g^* \rightarrow g^*g^*$, where star means non-zero invariant mass),
described by the the Lipatov-Altarelli-Parisi (LAP)  branching functions.
However, this approach is limited
in general, because a virtual gluon is not a {\it gauge invariant concept}.
In fact, LAP  approach  can  be used $only$ for soft gluons and for
radiation at {\it small-angle}: the term which is picked up is
the  leading power of the log.

 However, the soft radiation is exactly
the process which is strongly affected by the plasma screening effects:
and when those are taken into account, the log is not large enough to
keep the leading term only. Moreover,
as we will show below, the leading phenomenon is
not sequential radiation but simultaneous production of several gluons
with comparable momenta and at large angles.

All perturbative approaches mentioned share the same
  general uncertainty, related to
 the so far uncertain {\it infrared cut off} $p_0$,
separating soft and perturbative physics.
So far, none of them has
not yet been able to derive its value theoretically, or even locate the
specific phenomena responsible for it. Furthermore,
it was phenomenologically determined for pp case, but for nuclear collisions
it is expected to be different.

\subsection{The multi-gluon processes}

  Ref. \cite{SX_multiplication} have introduced
 {\it  the multi-parton processes}
 as a substitute for sequential branching.
  The specific problem addressed in that work is
gluon $chemical$
equilibration, while  $kinetic$ equilibration
 assumed.
  In this work the averaged matrix
elements for the
 multigluon processes was used, which are known for the sum of $tree$ diagrams.
 They
are defined {\it on mass shell} and therefore they
 are manifestly gauge invariant.
Furthermore, one can
separate the 'short-time' processes (for which a cascade approximation is
justified) from the 'long-time' ones  (such as interaction with collective
soft modes) by cutting off certain kinematical regions, in which none of the
kinematical invariants is small. The role of these processes at first impact
was discussed in the recent work\cite{multiglue} (see below).

  In this section we present some details about the multi-gluon QCD processes
on which the previous estimate was based.
 This discussion is limited to
gluons only, due to the following two reasons.
First,
the gluons do dominate
the nucleon structure functions at small x, as
well as scattering or production cross sections we are dealing with.
The second reason: matrix elements for
higher order gluon multiplication processes $gg\rightarrow
(n-2)g$  were clarified in the last few years only
and similar general expressions for quarks are still unknown.

  The so called ``Parke-Taylor formula'' \cite{Parke_Taylor}
to be used below is proven to be exact for
 n-gluon
processes in
the {\it maximum helicity violation} case. The squared matrix elements is
\be
 |M^{PT}_n|^2 = g_s^{2n-4} {N_c^{n-2}\over  N_c^2-1} \sum_{i>j}s^4_{ij}
\sum_{\rm P} {1 \over s_{12}s_{23}... s_{n1}}  \label{eq:pt}
\ee
In the above $s_{ij}= (p_i+p_j)^2$, the summation P is over
the $(n-1)!/2$ non-cyclic permutation of $(1...n)$.

   Unfortunately, the
exact result for other chiral amplitudes remains unknown.
However,  assuming that they are of
the same  magnitude  as the ``Parke-Taylor'' one, one gets some estimate
for the n-gluon matrix element. This was proposed by
Kunszt and Stirling \cite{Kunszt_Stirling} who add the following
 factor in front
of the ``Parke-Taylor'' formula
\be
 |M^{KS}_n|^2= KS(n) |M_{PT}|^2, {\rm \ \ with\ \ }
KS(n) ={2^n-2(n+1) \over n(n-1) } \label{eq:ks}
\ee

It agrees with the exact results for
$n=4$ and $n=5$, while
for higher orders
 a number of authors have checked this expression
up to $n=10$ using the Monte-Carlo generators, evaluating diagrams
numerically.
They have found that Eq.(\ref{eq:ks}) does a very reasonable job.

  Evaluation of the total cross section is a matter
 of integration over the many-body
phase space, which is  difficult to do analytically.
For symmetry, we introduce the universal cut off parameter $s_0$,
in the following way: $all$
 binary invariants are subject to a condition
\be
s_{ij}=(p_i+p_j)^2\ge s_0 \label{eq:constraint}
\ee,
 including all incoming and outgoing particles. This condition corresponds to
production of  the $resolved$ (=nonoverlapping)
jets.

 The {\it exclusive cross sections}
 should have the general form
\be
\sigma_n (s) = {1\over s_0} f_n( {s_0\over s} )
\ee

The functions $f_n$ are quite complex. It is analytically known
in the case of $n=4$
\be
f_4(\epsilon) = {9\pi\alpha_s^2\over 2} [ 1+{17 \epsilon \over 12}
-3 \epsilon ^2 +{ \epsilon^3\over 2}
-{\epsilon^4\over 3} -{\epsilon \over 1-\epsilon }
- \epsilon \log{ {1- \epsilon \over \epsilon }} ].
\ee
In the limit of  $\epsilon=s_0/s \rightarrow 0$, it tends to a constant.
For $n>4$, the functions $f_n$ are
 the polynomials of the $log(s/s_0)$, since each
binary invariant happen to be present in denominator only {\it once.}
The leading term of the total cross section should have the double log
behavior $ f_n(\epsilon) \sim [log^2(\epsilon)]^{n-4} $.
This can be best shown in  the soft-gluon case, where
the Parke-Taylor matrix element can be factorized as
\be
 |{\cal M}^{PT}_n|^2  \approx
( n-1 ) g_s^2 N_c
{1\over p_n^2 ( 1- \cos \theta ) } |{\cal M}^{PT}_{n-1}|^2 ,
\ee
where $p_n$ is the three momentum of the n-th gluon, $\theta $ is its
orientation with respect to any  one among the $n-1$ gluons.
The total cross section then has the form
\be
\sigma_n( \sqrt s )  \approx
 { n-1\over n-2}  {\alpha_s N_c \over 4\pi }
\int_{s_0} ^{s-s_0} {dM^2\over s-M^2} \sigma_{n-1} (M)
\int { d\cos\theta\over 1-\cos\theta }
\ee
When one proceeds iteratively, it is still true that each next particle
gives an
extra double log, so the answer should look as
\be
 \sigma_{gg\rightarrow (n-2)g} \approx \sigma_{gg\rightarrow gg}
[ \alpha_s N_c  C_n log^2(s/s_0)]^{n-4} \label{eq:sigman}
\ee
The  coefficient $C_n$ is however non-trivial to estimate.
Under a series of  approximations, it was shown to converge\footnote{
One may wander how square root of 3 can appear in expression for Feynman
diagram: the answer is there are $different$ coefficients for even and odd
n, both without such roots: but their geometric average has it.} to
\cite{Goldberg_Rosenfeld}
\be C_n \rightarrow {1 \over 4\pi\sqrt{3}} \ee

Note that there is no factorial suppression of large n: it originated from
constructive interference of n! ``strings" in the $squared$ matrix element.
Therefore, instead of exponential series, one has in fact a $geometric$ one.
This asymptotic behavior signals a warning to the eligibility of
perturbative theory since squared log can overcome the coupling constant and
the $total$ cross section $\sigma_{tot}=\Sigma_n \sigma_n$
 would diverge. We will return to this problem in the section, devoted to SSC
energies.

The  energy dependence of the
exclusive cross sections was obtained numerically,
and it can be parameterized
by the following analytic expressions
\be S_0\sigma_n(M)\  [{\rm GeV}^2 mb ] = (\alpha_s/\alpha_{s0})^{n-2}
10^{ a_n + b_n (\log_{10}s {2M\over  \sqrt{s_0} } )^{c_n} }.\ee
The parameters are found to be
\be
a_5,\ b_5,\ c_5 &=& 1.0175,\ -1.6675,\ -1.977 \\
a_6,\ b_6,\ c_6 &=& 2.1323,\ -3.6323,\ -1.688 \\
a_7,\ b_7,\ c_7 &=& 2.8426,\ -5.6426,\ -1.871
\ee
One can see that large n processes are more sensitive to
$s/s_0$, and they also are more important for larger $s/s_0$.

The main point is that the leading log approximation
is not reliable for the problems related with the minijets, because
the typical $s$ is only several times larger than $s_0$; so
picking only the leading
$\log^{2n} (s/s_0) $ terms is not justified. Therefore, we
study the parton multiplication processes
considering all the kinematic regions and interference effects.

  Now we move from cross sections to one-body (exclusive) distributions. We
have found very peculiar consequences of the
  Parke-Taylor formula, which is
   significantly different  from the picture one is used to in QED radiative
processes. This
is demonstrated in Fig.17
where we show the transverse momentum $p_t$ and
rapidity y distributions of secondary gluons,
for multigluon processes ( $gg\rightarrow (n-2)g$ ). We have chosen
a jet
 resolution $s_0/s = 0.02^2 $.

Going from n=4 (elastic process) to larger n
one can see that the particle distribution begin to build up very rapidly at
central rapidity.
When $n=5$, a  soft gluon radiation
 is filling the gap between the  two major
outgoing gluons: the rapidity distribution becomes flat.
This result is well known: not that it already significantly deviates from the
 QED case, where there exist a dip at mid-rapidity caused by destructive
interference of radiation in initial and final states.

 When n is increase further, all the outgoing particles
are piled up around $y=0$.
Its width is O(1), so the angular distribution is in fact nearly isotropic.
This can be traced to $constructive$ interference of many diagrams: soft gluons
are effectively emitting each other.

Let us now proceed to the $p_t$
spectrum. It is somewhat surprising to see, that for larger n
it becomes roughly exponential,
in the large range of $p_t$. Moreover,
the slope is almost $universal$ for all n processes, about 1\% of the total
energy.
 Thus, something like thermal\footnote{
One should not  confuse this phenomenon neither
with true thermalization of
partons, for which their rescatterings are needed, nor with exponential
$p_t$ spectrum observed in pp collisions due to soft hadronic processes.}
 $p_t$ distribution of gluons
is produced already in $one$ multigluon scattering event!

   These distribution can be compared with the leading log or
'soft gluon approximations',
predicting {\it flat} rapidity distribution $d\omega/\omega=dy$ and {\it
power like} $p_t$ spectra $d p^2_t/p^2_t$. Clearly, such approximation is
qualitatively wrong for the kinematical region under investigation.

	The inclusive cross section for n-jet
 production from hadronic collision can
be calculated from convoluting the n-gluon matrix element with
the luminosity function
\be
E_1E_2...E_{n-2}{d\sigma_n \over d^3p_1d^3p_2...d^3_{n-2} }
 =\int dx Lum(x,s_0) {1\over 2 x^2 s } |{\cal M}_n|^2
(2\pi)^4 \delta^4 ( \sqrt s - \sum_{i=1}^{n-2} p_i )
\label{eq:jet}
\ee
which is the probability that the initial two partons
carries x fraction of the total invariant mass and depends on the
gluon structure function as
\be
Lum(x,Q^2)= \int_x^{\sqrt{x}} dx_1 G(x_1, Q^2) G(x^2/x_1, Q^2) {2x\over x_1}
\ee
We define a number
\be
\delta \equiv \sqrt{ s_0/s }
\ee
to be the ratio of the cutoff mass to the total (center of mass)
\setcounter{footnote}{0}
 energy\footnote{
Note that in $\delta$ it is the total energy of collided hadrons, not partons,
as for the parameter $\epsilon$ considered above. }
 and
will use it for analysis later on.
In this notation HIJING's cut-off is $\delta= 0.014$.

The inclusive jet production in pp collisions
was studied in multiple experiments: those can be well reproduced by
 gluon jets from gg-gg only. One should therefore consider $exclusive$
multi-jet events, to test these formulae.
Furthermore,
   formulae discussed above correspond to the tree-level diagrams, so
the natural question is what higher-order corrections may do to them.
Experience with binary processes suggests that
 even in
the kinematic region where they should work, one gets the so called K-factors,
changing the cross section by a factor 2 or so. For multi-gluon processes,
with many kinematical variables, the radiative corrections are generally
a very complicated functions of many of them, leading presumably to some sort
of
formfactors.

   In \cite{multiglue} we have used data on  multijet measurement are
obtained by the UA2 collaboration \cite{UA2} at CERN
SPS collider energy $\sqrt s= 630 GeV \bar p p$ collision.
The data on 4 jet production have especially good statistics
and each of the four jets
are required to have $p_t$ larger than 15 GeV, it
corresponds to $x_{min} \approx 0.05$ \footnote{
Note that this energy is factor 3.1
higher than the nominal RHIC one, therefore scaling by this factor down one
gets ``mini-jets" of about 5 GeV, which is not very far from those we
consider in relation with QGP thermalization at RHIC.}
   The first level of our investigation was studies of the transverse
momentum distributions for different jet number.
It actually agree with
 the universal exponent found in the previous section.
The slope furthermore is indeed almost the same  for different n's.
 Thus, whatever $K-factor$ may be, it is
presumably $not$
a strong function of $p_t$, but more or less constant in the whole
kinematical domain.
  The second level is the $absolute$ values for exclusive
cross sections with different n, which is suppose to tell us what
the magnitude of this K-factor might be.
The authors of \cite{UA2} themselves
 have compared their data with the  exact
matrix elements calculation \cite{BGK},
and with  the
``improved" Parke-Taylor formula in \cite{KS86}.
They have reported agreements with data within an impressive 20\%:
so, this formula really works!

    $Assuming$ kinetic equilibrium of gluons,
or momentum distribution  $ f( p_t ) = \xi \exp ( -p_t/ T  ) $ at  time
$\tau_{kinetic}\sim 0.3 fm$, one can consider gluon multiplication leading to
$chemical$ equilibration of glue, at which $\xi\rightarrow 1$. Time evolution
of gluon fugacity and temperature \cite{SX_multiplication}
 is shown in Fig.18 for 3
scenarios, with intial values $(T,\xi)=(0.56,006),(0.5,0.25),(0.5,0.5)$.
The dashed curve is for gg into ggg only, while two others include multigluon
processes in two different approximations. One may conclude from these results,
that chemical equilibration of gluons proceed sufficiently rapid, and is
concluded  during the lifetime of QGP.

   Let us now proceed to initial impact in AA collisions.
As it was explained above,
already after the first multi-gluon scattering the spectrum
has momentum distribution of the type $ f( p_t ) = \xi \exp ( -p_t/ T  ) $.
The slope $T$ is not very sensitive to the cut-off, for RHIC energy
$T= 2$ GeV.
But the fugacity (and the total entropy)
 is strikingly sensitive to the cut-off.

 It is important
that in AA central collisions the situation is completely different
from the pp case: partons are
in dense system of their neighbours.
If anything, the lessons from finite-T QCD is that parton
interaction in dense systems leads to $density-dependent$
screening, which makes the direct extrapolation from pp case impossible.
  Some guidance can probably be obtained from the equilibrium situation, for
which
lattice data  exist. For $T > 2-3
T_c$ they tell us that $M_{eff}\approx  2 T$. If applied to gluon system
$after$ equilibration, with $T\approx 500 MeV$, one gets
$M_{eff}\approx$ 1 GeV. It can probably be taken as some lower bound
of the cut-off momenta.

The physical cutoff may be the screening mass of the parton system.
In a non-equilibrated  plasma,
one expects that
\begin{equation}
 m^2\approx g^2_s \xi T^2
\end{equation}
So the cut-off should be larger for higher energy hadronic collisions,
because the partonic system is denser.
We have calculated the number of produced gluons
using the self-consistent cut-off: it leads to
initial temperature and fugacity at RHIC $T\approx 2 GeV,\xi=  0.1 $.
If so, the produced entropy is on the large side, roughly comparable to that
predicted by PCM.
The corresponding time scale is just time of the first collision, by
uncertainty relation it is about .1 fm/c .

\subsection{The LHC energies: limitations of the
perturbation theory}

   We have discuss the LHC case in a separate subsection, because here we seem
to find a serious problem. But before we come to it,
  let us briefly consider the mini-jet
production at LHC in general.

Kinematically, in this case one is dealing with x$\sim 10^{-4}$, and
 we now know from
HERA measurements  that nucleon gluonic structure functions experience strong
grouth in this region\footnote{
Whether $nuclear$ structure function also grow,
as the $nucleon$ does remains a matter of controversy.
For heavy ions at LHC
one may finally find the so called
$saturation$. }.
 In recent paper \cite{EKR} those were taken into
account, and with $binary$ processes  with the HIJING cutoff they
have  evaluated the number of mini-jets produced by binary scattering.
The result
suggests nearly chemically equilibrated
 system of gluons at the time $\tau=1/p_0=0.1 fm/c$, with $T\approx 1 GeV$.

  Unfortunately,
this calculation is completely destroyed
\cite{multiglue} if one includes
the multi-parton processes. Each subsequent subprocess
$2\rightarrow (n-2)$ lead to the estimate $larger$ than (n-1)-st!
The numbers we get are 0.35, 1.32, 3.35, 15.11,
for n=4,5,6, and 7 at $\delta = 0.00045$, which
corresponds to the HIJING cut-off $p_0 = 2$ GeV.
The effect calculated in ref.\cite{EKR} is therefore just a beginning of a
divergent\footnote{Mathematically, the series
are of course limited by the applied cut off, so they are not really
divergent. However, one can hardly take
the scenario of rapid transformation of TeV gluons into thousand GeV ones
for an answer.} geometrical series!

   The reason this happens is that
 we are actually {\it beyond applicability limits of the
perturbation theory}. Although $\alpha_s << 1$, the powers of log($s/s_0$)
has overcome it.
The only comment we may make here is that
probably in so dense partonic system the cut off should be much larger. We have
calculated and shown the number of gluons produced at LHC in the
Table 1 examples, for few values of $p_0$. Thus, either (i) there exist some
mechanism producing high enough cutoff $p_0\sim 10 GeV$ and the perturbation
theory is then justified, in its reduced domain; or (ii)
has to develop and apply some non-perturbative methods even for multi-GeV
gluon jets.


\section{Summary}

   The main lesson from the first part of this paper is as follows: instantons
are the dominant dynamical phenomenon, as far as physics of light quarks and
lowest hadrons is concerned. Random liquid and quenched lattice data give
very consistent results for correlators
in many channels, for various wave functions etc. Glueballs also can be
qualitatively related to instantons, and one can see there direct
evidences for strong
classical fields and self-duality.
The unquenched results for the instanton approach are under way, as well as
their lattice analogs. One effect considered, the screening of the topological
charge, is especially interesting, and it should be studied in much greater
details.

  The second part deals with various consequences of an idea,
attributing the chiral phase transition in QCD to
 a rearrangement of the instanton liquid, going from a {\it random
phase} (at low T) to a correlated phase of {\it polarized
$\bar I I$ molecules} at $T>T_c$. In this
scenario,  a significant number of  instantons is present
 at temperatures $T=(1-2)T_c$, causing a variety of
nonperturbative effects.

One of them is {\it polarization} of
$\bar I I$ molecules.
at temperatures $T\simeq T_c$, related with
 their {\it significant
contribution to the energy density and pressure} of the system
near the phase transition region.
    The presence of $\bar I I$ molecules above $T_c$
also produces quite {\it specific interactions between light quarks}, which is
 $U(2)\times U(2)$ symmetric.
 This results is in agreement with lattice
simulations, in which the presence of an $attractive$ interaction in
the scalar channel (but not vector ones)
 has been established from an analysis of
spacelike screening masses.

Finally, we have discussed a recently proposed strategy for the
experimental search for
the QCD phase transition.  The usual view is that the QGP will reveal
itself as we go to higher and higher energy density, since then the signals
from QGP should outshine that of the hadronic background.  We certainly
have no dispute with this approach.  However, it seems also possible
to go $down$ in energy from the nominal SPS one,
looking for the ``softest point", at which
   evolution of the excited
matter leads to
especially
long-lived fireball.

In the last chapter
 we have studied QGP properties, as it is expected to be at RHIC and LHC
energies. Our main concern was the proper language which is needed to describe
``parton cascades". We have argued that it is
the multi-gluon processes gg $\rightarrow$(n-2)g, which can be done
using improved Parke-Taylor formula. First of all,
we have found some new features
of these processes, such as (i) piling up at mid-rapidity, and
(ii) $exponential$ $p_t$ spectrum of gluons, with nearly universal
slope.

   Using those processes, we have found that gluonic
component of the plasma can be chemically
equilibrated during its lifetime: but quarks cannot.
   Proceeding to the first impact in AA collisions,
 we have found that here the multi-gluon processes (n>4)
 are much more important than lower-order ones. At RHIC energies
 we have proposed a self-consistent evaluation of initial conditions for a
cascade.
At LHC, the situation is more dramatic and (in contrast to what was reported
earlier)
perturbative predictions start to diverge, in the
sense that each next process is more probable than the previous one.

\newpage
\begin{figure}[t]
   \vspace{4.5in}
   \caption{ Transverse momentum and rapidity distribution of gluons produced
in multi-gluon processes with different number of participating gluons, from
n=4 (elastic scattering) to n=8. }
\end{figure}

\begin{figure}[t]
   \vspace{4.5in}
   \caption{Time dependence for temperature and fugacity during
 ``chemical equilibration" of gluons: see text for explanations. }
\end{figure}

\end{document}